\documentclass[twocolumn]{aastex7}

\newcommand\gva{{\sc Genec}}

\begin{document}

\title{Growth of Metal-Enriched Supermassive Stars by Accretion and Collisions}

\author[orcid=0000-0003-1927-4397,sname='Devesh Nandal']{Devesh Nandal}
\affiliation{Center for Astrophysics, Harvard and Smithsonian, 60 Garden St, Cambridge, MA 02138, USA}
\email[show]{deveshnandal@yahoo.com}  

\author[orcid=0000-0003-4330-287X,gname=Bosque, sname='Sur America']{Sunmyon Chon} 

\affiliation{Max-Planck-Institut f$\ddot{u}$r Astrophysik, Karl-Schwarzschild-Str. 1, D-85741 Garching, Germany}
\email{sunmyon@MPA-Garching.MPG.DE}

\begin{abstract}

Supermassive stars (SMSs) are candidate progenitors of massive black hole seeds and may contribute to anomalous abundance patterns in high-redshift galaxies and globular clusters. Recent radiation-hydrodynamic simulations indicate that SMSs can form at finite metallicity, not only in metal-free direct-collapse conditions. We model SMS growth with \textsc{GENEC} over $Z/Z_\odot=10^{-5}$--$10^{-2}$ using simulation-motivated accretion histories. The final masses reach $\sim7.2\times10^{4}\,M_\odot$ at $10^{-5}\,Z_\odot$ and $\sim2.3\times10^{3}\,M_\odot$ at $10^{-2}\,Z_\odot$. Models are evolved through the pre-main sequence and core H-burning phases, terminating at the onset of general-relativistic instability for $Z\lesssim10^{-4}\,Z_\odot$ or at core He exhaustion for $Z\gtrsim10^{-3}\,Z_\odot$. The dominant mass growth channel transitions from collision-driven to accretion-driven between $Z=10^{-4}$ and $10^{-3}$. With stellar lifetimes remaining nearly constant at $1.8$--$2.0$ Myr, collisions do not significantly rejuvenate the star, implying that collision driven runaway collapse cannot proceed in isolation and must be supplemented, and likely dominated by gas accretion. We further compute the critical inflow rate required to keep the stellar envelope inflated, $\dot{M}_{\rm crit}$, which decreases with increasing $Z$ and decreasing central mass fraction of hydrogen ($X_{\rm c}$). The critical rate falls below $10^{-5}\,M_\odot\,{\rm yr^{-1}}$ at $X_{\rm c}=0.60$ for $10^{-2}Z_\odot$. This indicates that SMSs with $0.01~Z_\odot$ are cool supergiants during most of their lifetimes, where UV photon emissivity and radiative feedback is strongly suppressed. Overall, SMS evolution remains viable up to $Z\simeq0.01\,Z_\odot$, supporting SMS formation in proto--globular clusters and other metal-enriched dense environments.

\end{abstract}


\section{Introduction}\label{sec:intro}

The discovery of luminous quasars at $z\gtrsim 6$ hosting $\sim10^{9}\,M_\odot$ black holes poses a formation timescale problem for standard light seeds with Eddington-limited growth \citep{Inayoshi2020,Volonteri2021}. Surveys now count hundreds of such objects, and JWST is extending the census to fainter Active Galactic Nuclei (AGN) and higher redshift \citep{Banados2016, Kocevski2023}. Two families of solutions dominate current models. One invokes massive seeds formed by direct gas collapse or SMS progenitors \citep{Loeb1994,Bromm2003, Begelman2006, Chon2018, Wise2019, Inayoshi2020}. The other relies on sustained episodes of super-Eddington accretion that compress the growth timescale \citep{Lupi2024,Bhowmick2024}. Recent empirical inferences suggest that both heavy seeds and intermittent super-Eddington phases may operate in the early Universe \citep{Andika2024,Maiolino2024}. 

In parallel, JWST spectroscopy has revealed galaxies with near- to super-solar N/O at sub-solar O/H, which standard yields struggle to reproduce \citep{Bunker2023,Cameron2023,Harikane2025, Isobe2025}. Proposed explanations include hot H-burning in very massive or supermassive stars, and alternative chemical-evolution pathways with bursty star formation and differential winds \citep{Charbonnel2023, Marques2023, Nagele2023,Nandal2024}. These data place joint constraints on early star formation, nucleosynthesis, and black-hole seeding.

SMSs provide a single mechanism that can address both problems simultaneously. On one hand, an SMS of mass \(\gtrsim10^{5}\,M_{\odot}\) collapsing yields a heavy seed that can accelerate SMBH growth \citep{herr23a,Latif_2014c}. On the other hand, the hot hydrogen burning cores of SMSs produce distinct nucleosynthetic signatures such as high N/O, low C/O and Ne/O that match anomalous abundances seen in early galaxies \citep{Nandal2025}. Thus SMSs offer a unified path connecting structure formation, seed black holes, and chemical fingerprints in the first galaxies.

The formation of SMSs has been intensely studied in the context of star formation in zero-metallicity environments.  
Stars that form from primordial, metal-free gas clouds are termed ``Population~III'' (Pop~III) stars, whose typical masses range from \(10\) to \(1000\,M_\odot\) \citep[e.g.,][]{Stacy2013, Hirano2014, Susa2014, Hirano2015, Jaura2022, Prole2022, Sugimura2023, Sharda2025}.  
Under extreme conditions, stellar masses can reach up to \(10^{5}\,M_\odot\), producing SMSs that eventually collapse into heavy seed black holes of comparable mass \citep{Shibata2002, Uchida2017, Fujibayashi2025}.  
This process is commonly referred to as the ``Direct Collapse (DC)'' model.  
A key physical condition for SMS formation is the high gas temperature or equivalently, high pressure of the collapsing cloud.  
In this model, strong far-ultraviolet (FUV) radiation suppresses molecular cooling keeping the cloud temperature at \(\sim10^{4}\,\mathrm{K}\) \citep{Omukai2001a, Inayoshi2014}. 
Such high temperatures drive rapid gas accretion onto the central protostar \citep{Whitworth1985}.  
Further multi-dimensional simulations have confirmed that the formation of SMSs in metal-free strong FUV environments \citep{Latif2016, Shlosman2016, Chon2018, Whalen2020, Matsukoba2021, Prole2024}.

Additional dynamical effects such as turbulence \citep{Latif_2022} and the streaming velocity between baryons and dark matter \citep{Hirano_2017, Schauer_2017} can delay collapse until the halo grows massive enough to accumulate sufficient gas, thereby enabling SMS formation.

Historically, a zero-metallicity condition was considered essential to prevent fragmentation by keeping the gas temperature high during collapse. 
Even trace amounts of metals or dust grains can induce efficient cooling, causing the temperature to drop abruptly from \(10^{4}\,\mathrm{K}\) to a few hundred \(\mathrm{K}\) \citep{Omukai2008}.  
This rapid cooling was thought to trigger vigorous fragmentation \citep{Li2003,Jappsen2005}, thereby suppressing SMS formation.  
However, recent studies have shown that massive gas accretion can still occur in environments with finite metallicity (\([Z/H] \lesssim -3\)), allowing SMS and heavy seed black hole formation even under such conditions \citep{Chon2020, Chon2025}.  
The fragmentation does occur to form a number of low-mass stars, while stellar collisions promote to form SMSs \citep{Schleicher2022, Reinoso2023, Solar2025}.
These works demonstrate that while fragmentation is enhanced in metal-enriched clouds, filamentary inflows and collisional mergers can still assemble central stars with masses of \(10^{4}\!-\!10^{5}\,M_\odot\).  
Moreover, simulations of gas discs formed during galaxy mergers have revealed the formation of metal-enriched supermassive cores with masses exceeding \(10^{4}\,M_\odot\), consistent with SMS formation \citep{Zwick2023}.  
Together, these findings suggest that SMSs need not be strictly Pop~III objects, but can also form in moderately metal-enriched environments such as proto–globular clusters or high-redshift compact starbursts \citep{Gieles2018}

In globular–cluster formation models, a central supermassive star can assemble by collisional runaway in a young, dense core \citep{Freitag2006, Fujii2024, Lahen2025,Chon2025, Vergara2025b,Vergara2025}. Such runaways are favored when the central relaxation time is short and mass segregation is rapid, which drives repeated mergers among massive stars \citep{Moeckel2011}. The SMS scenario was proposed to explain the light–element anomalies and He spreads in present–day clusters through hot–H–burning yields and subsequent dilution \citep{Denissenkov2014}. These models link nucleosynthetic patterns (C–N, Na–O, Mg–Al) to processed ejecta from a $\gtrsim10^{3}$–$10^{4}\,M_\odot$ object and predict strong N enrichment at low metallicity, consistent with recent high–$z$ inferences \citep{Charbonnel2023}. The feasibility of sustained growth depends on the competition between collision–driven mass gain, gas inflow, and winds, with metallicity and $L/M$ setting the outflow regime \citep{Glebbeek2009}. Thus SMSs provide a common framework for multiple populations in clusters and for rapid heavy–seed formation in gas–rich environments, while placing testable constraints on dynamics, feedback, and chemical evolution \citep{Zwick2023}.

In this work we model metal-enriched supermassive stars with metallicities \(Z/Z_\odot=10^{-5}\)–\(10^{-2}\) using time-dependent accretion histories from radiation-hydrodynamic cluster-collapse simulations \citep{Chon2025}.  Section \ref{sec:2} describes the stellar models and accretion treatment, Section \ref{sec:results} examines the overall evolution, explores the impact of variable accretion histories, evolution of the critical accretion rate during core-hydrogen burning, and quantifies rejuvenation from collisional mass gain. We then compare our results with other works and provide a first order estimates on mass loss in Section \ref{sec:discussion}. Finally, in Section\ref{sec:conclusion} we summarise our findings and discuss the future work.  
  
\section{Methods}\label{sec:2}

We outline here the numerical framework and physical assumptions adopted for modeling metal-enriched supermassive stars using \gva \citep{Eggenberger2008, Nandal2024e, Nandal2025} and the large-scale cluster formation which gives the mass accretion histories assumed in our model \citep{Chon2025}. The subsections that follow describe the stellar models, accretion histories, and code modifications in detail.

\subsection{Stellar models and initial conditions}

Four baseline models span metallicities $Z/Z_\odot=\{10^{-5},10^{-4},10^{-3},10^{-2}\}$ and share identical numerical settings. Additional sequences extend the grid: sixteen variable-accretion models at $Z=10^{-4}\,Z_\odot$ used to examine the critical accretion rate during core-H burning, a constant-accretion model at $Z=10^{-2}\,Z_\odot$ to explore the impact of variable and constant accretion histories, and two $Z=10^{-4}\,Z_\odot$ tests incorporating different empirical wind prescriptions.  Table~\ref{tab:grid} summarises the computed models, their compositions, and terminal evolutionary states.  The final phase corresponds to the onset of general-relativistic (GR) instability for $Z\!\le\!10^{-4}$ and to core-He burning for the higher-$Z$ cases.  The GR instability is identified through linear adiabatic pulsational analysis following \citet{Saio2024} and \citet{Nandal2024d}.  All models are treated as cold accreting; the accretion luminosity is radiated away, and no entropy is advected into the stellar interior. This cold boundary condition is an idealised limit that assumes efficient radiative cooling of the accretion shock and the inner flow. It may be violated during the most intense and short-lived bursts present in the imported histories, when some fraction of the accretion energy could be retained and raise the entropy of the outer envelope. Such ‘hot’ accretion would be expected to primarily shift the normalisation of the radius response and of any inferred \(\dot M_{\rm crit}\), but it does not relax the fact that the time available for growth during core-H burning is bounded by the nuclear clock. Two exploratory models at $Z=10^{-4}\,Z_\odot$ incorporate mass-loss rates from the \citet{DeJager1988}, \citet{Vink2001}, and \citet{Kudritzki2002}.  These runs are designed solely to map the parameter space of possible line-driven winds in metal-enriched SMS envelopes.

\begin{table*}[t]
\centering
\caption{Model grid and adopted compositions. Each model starts from a 10\,M$_\odot$ fully convective seed.}
\label{tab:grid}
\begin{tabular}{lcccccccc}
\hline
Metallicity & Accretion & Final stage & Comments & $X_{\rm ini}$ & $Y_{\rm ini}$ & $Z_{\rm ini}$ & $M_{\rm final}$ (M$_\odot$) & $t_{\rm final}$ (Myr) \\
\hline
$10^{-5}\,Z_\odot$ & Variable & GR inst. & Baseline model & 0.751600 & 0.248400 & $1.4\times 10^{-7}$ & 72492 & 1.948 \\
$10^{-4}\,Z_\odot$ & Variable & GR inst. & Baseline model & 0.751597 & 0.248402 & $1.4\times 10^{-6}$ & 82193 & 1.962 \\
$10^{-4}\,Z_\odot$ & Variable & GR inst. & $\dot{M}_{\rm crit}$ test & 0.751597 & 0.248402 & $1.4\times 10^{-6}$ & -- & -- \\
$10^{-4}\,Z_\odot$ & Variable & GR inst. & Wind pres. & 0.751597 & 0.248402 & $1.4\times 10^{-6}$ & -- & -- \\
$10^{-3}\,Z_\odot$ & Variable & Core He-b & Baseline model & 0.751568 & 0.248418 & $1.4\times 10^{-5}$ & 32629 & 2.029 \\
$10^{-2}\,Z_\odot$ & Variable & Core He-b & Baseline model & 0.751284 & 0.248576 & $1.4\times 10^{-4}$ & 2270 & 1.829 \\
$10^{-2}\,Z_\odot$ & Constant & Core He-b & Time Avg. $\dot{M}_{\rm acc}$ & 0.751284 & 0.248576 & $1.4\times 10^{-4}$ & -- & -- \\
\hline
\end{tabular}
\end{table*}

\subsection{Accretion and collision histories}

The time-dependent accretion histories are taken from the hydrodynamical simulations of star cluster formation with SMSs presented in \citet{Chon2025}.  
The sample of gas clouds is extracted from the cosmological simulations with a box size of \(20~h^{-1}\mathrm{Mpc}^{-3}\) described in \citet{Chon2016}, where $h=0.677$ is the dimensionless Hubble constant \citep{Planck2014}.  
In that study, semianalytical modeling was combined with \(N\)-body calculations using \textsc{Gadget2} \citep{Springel2005}, identifying pairs of luminous source galaxies and chemically pristine atomic-cooling halos as candidate sites for SMS formation.  
Subsequent radiation hydrodynamical simulations with \textsc{Gadget3} \citep{Springel2005} confirmed that two of these candidate halos indeed form SMSs.  
Molecular cooling is completely suppressed by strong FUV radiation from the nearby source galaxy, maintaining the gas temperature at \(\sim10^{4}\,\mathrm{K}\).  
After the formation of a protostar, sustained massive gas accretion enables efficient mass growth, leading to the formation of an SMS within the cloud \citep{Chon2018}.  

\citet{Chon2025} extended the calculations to \(2\,\mathrm{Myr}\) after the initial protostar formation and varied the cloud metallicity over the range \(-6 < [Z/H] < -2\).  
Their simulations include a non-equilibrium primordial chemical network, detailed cooling and heating processes, and photo-heating effects from nearby stars through photo-ionization, photodissociation of molecules, and dust irradiation.  
The accretion rate onto the protostars was measured using a sink particle method, in which gas particles within \(1000~\mathrm{au}\) of the sink are accreted to increase the protostellar mass.  
The sink particles are assumed to merge when a pair of them approaches within a distance equal to the sum of their sink radii and becomes gravitationally bound to each other.

Gas flows toward the cluster center through filamentary streams and stellar encounters, a process referred to as super-competitive accretion.  
At \(Z \lesssim 10^{-3}\,Z_\odot\), this mechanism builds central SMSs exceeding \(10^{4}\,M_\odot\); at higher metallicities, enhanced dust and metal-line cooling induce fragmentation, producing compact star clusters dominated by very massive stars (\(\sim10^{3}\,M_\odot\)).  
The simulation outputs provide the net mass-delivery rate to the central object every \(500\,\mathrm{yr}\), a cadence chosen by the typical dynamical timescale in the sink region.  
Each recorded \(\dot{M}\) value incorporates both smooth inflow and episodic collisional events within the forming cluster, thereby capturing transient bursts of accretion. The cold-accretion assumption is therefore most uncertain during the largest spikes, where partial trapping of accretion energy could occur. These accretion histories are imported into \gva, ensuring that both accretion- and merger-driven growth are treated self-consistently up to the end of the main sequence, when inflow ceases in the 3D models.  

Note that the stellar evolution model adopted in \citet{Chon2025} is relatively simple, interpolating the pre-calculated stellar structure table \citep{Hosokawa2009}. 
The stars are assumed to evolve in two distinct phases depending on the accretion history: the supergiant-star phase and the main-sequence phase.  
When the accretion rate exceeds the critical value of \(0.02~M_\odot~\mathrm{yr^{-1}}\), the star is assumed to enter the supergiant-star phase.  
In this phase, the stellar effective temperature is fixed at \(6000~\mathrm{K}\), and the luminosity is set to the Eddington limit.  
Once the accretion rate drops below the critical value, the star gradually transitions to the main-sequence phase on a timescale given by the surface Kelvin–Helmholtz time \citep{Sakurai2015}.

\subsection{Numerical implementation}

The pre-MS module of \gva\ was expanded to allow accretion of finite-metallicity elements. The surface composition of accreted gas is fixed to the initial $(X,Y,Z)$ values specified for each model and remains constant through the evolution. A dedicated time-step controller monitors instantaneous $\dot{M}$ and restricts the added mass per step to maintain numerical stability, even when the inflow changes by several orders of magnitude between adjacent 500 yr records.  This mechanism prevents sudden jumps in luminosity and structure during rapid bursts.  The modifications build upon earlier numerical refinements in \citet{Nandal2023,Nandal2025b}.  Equation-of-state and opacity treatments follow OPAL tables, using the same interpolation and coupling procedures as in our zero-metallicity SMS studies \cite{Ekstrom2012, Nandal2024b}.

\subsection{Treatment of collisions}

Collisions are represented as rate spikes within the imported accretion histories rather than as explicit hydrodynamic mergers.  No additional heating or entropy deposition is imposed; instead, the total inflow rate is simply enhanced over the sampling interval corresponding to a collision event. This assumption effectively provides an upper limit to the collisional mass gain by assuming complete mass retention and neglecting radiative losses. The method isolates the maximum mass achievable through mergers under idealised conditions. In Genec the stellar response is computed only for the total imposed mass-delivery history \(\dot M(t)\) from the simulations. We therefore use the terms \emph{accretion} and \emph{collision} below as bookkeeping labels for the smooth baseline and the impulsive spikes in \(\dot M(t)\), rather than as distinct internal-physics modes. In particular, phrases such as “collision-dominated growth” simply denote epochs where the adopted \(\dot M(t)\) history is dominated by impulsive spikes, while the stellar-evolution boundary condition remains the same in all cases. Taken together with cold accretion, this setup is deliberately optimistic for collision-assisted growth because it maximises mass growth and retention while minimising additional thermal energy deposition during burst episodes. A more realistic treatment could include impact heating and partial trapping of accretion energy, which would tend to make envelope inflation easier during bursts, but it would not extend the evolutionary time available for assembling the final mass. We therefore interpret our collision-only growth tests as best-case upper limits on what can be achieved within a main-sequence lifetime.

\section{Results}\label{sec:results}
\begin{figure}
    \centering
    \includegraphics[width=\linewidth]{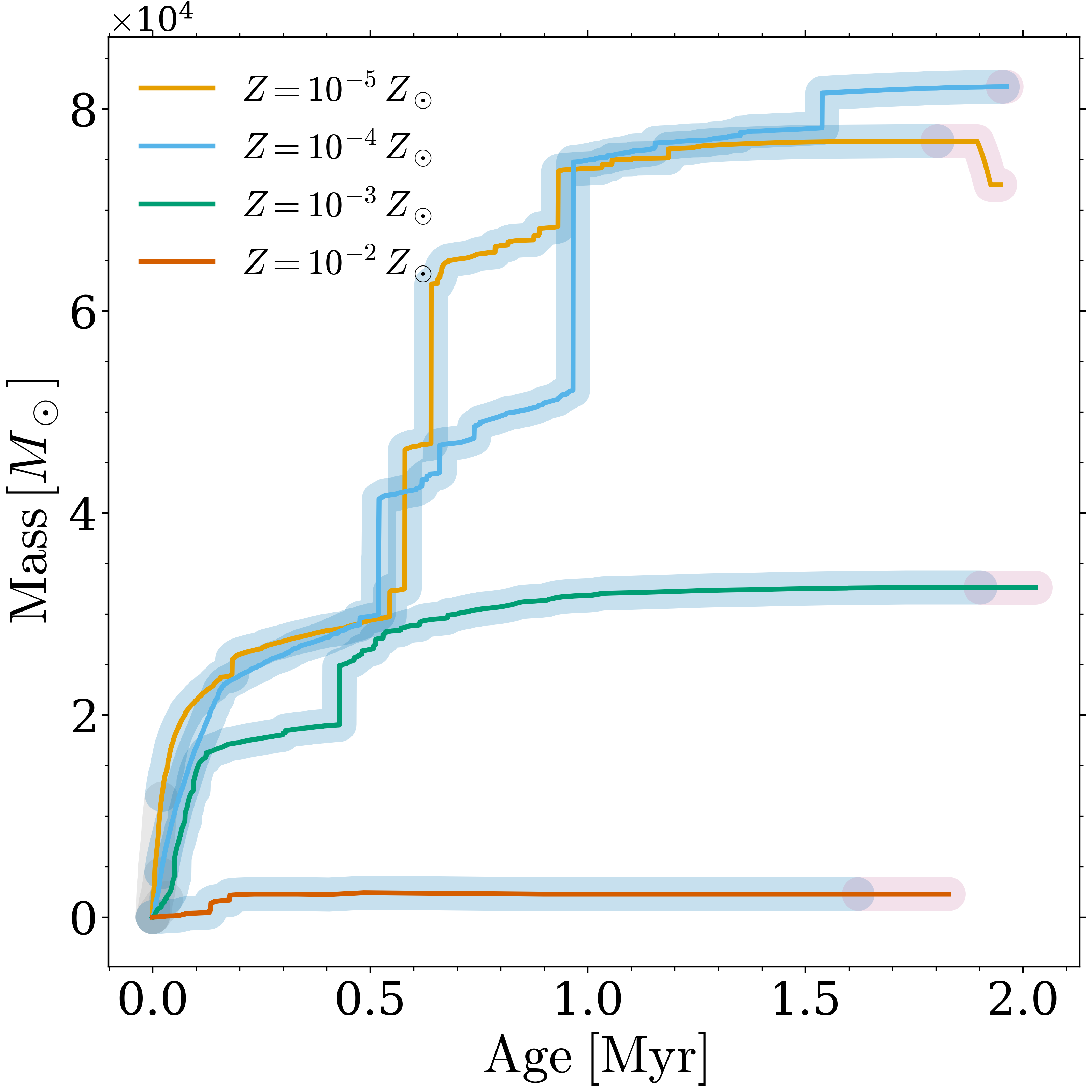}
    \caption{Mass versus age evoluton of models at four metallicities. Colored lines denote the metallicities assumed in the models with $Z/Z_\odot=10^{-5}, 10^{-4}, 10^{-3}, 10^{-2}$ as labeled. Phase underlays follow each track: gray pre--MS ($X_{\rm c}\ge0.74$), blue MS ($0.01\le X_{\rm c}<0.74$), magenta post--MS ($X_{\rm c}<0.01$).}

    \label{fig:3_new_MvsT}
\end{figure}

We begin this section by exploring how mass growth shapes the overall evolution of our stellar models. Figure~\ref{fig:3_new_MvsT} shows the stellar mass as a function of time for the four models. The stars gain mass through gas accretion and stellar collisions: mass growth starts during the pre-main-sequence phase (light grey) and continues until the end of core hydrogen burning (light blue).
All models show a similar behavior: they assemble about 90\% of their final mass within the first $\sim$1 Myr, after which the mass growth slows down. This slowdown occurs because the gas reservoir becomes depleted around this time. Beyond $\sim$1 Myr, the mass increase is therefore dominated by discrete collisional events rather than by smooth accretion. We return later to the individual evolutionary tracks and discuss in detail when and how this transition occurs in each case.
Across the four metallicities, the final stellar mass decreases monotonically with increasing metal content. This is because more efficient cooling reduces the initial Jeans mass and thereby lowers the gas accretion rate. The total lifetime shows no clear dependence on metallicity. Although collisions can occasionally supply fresh hydrogen to the core and partially rejuvenate the star, the accompanying increase in the hydrogen-burning rate at higher mass largely compensates for this effect, so the net impact on the lifetime is small. These effects, and their influence on stellar lifetimes, are examined in detail in the following section.

We briefly summarize the mass-growth histories across metallicities, focusing on three common phases:
(i) the early build-up within the first $10^{4}\,\mathrm{yr}$,
(ii) the subsequent evolution up to $\sim 1\,\mathrm{Myr}$ while gas is still available, and
(iii) the late-time, merger-dominated growth after gas accretion is shut down.

For $Z/Z_\odot \lesssim 10^{-4}$, the stellar mass rises rapidly to $\sim10^{4}\,M_\odot$ within the first $10^{4}\,\mathrm{yr}$, mainly through gas accretion and mergers with low-mass stars ($<100\,M_\odot$). The accretion rate then gradually declines as the gas supply from the outer envelope diminishes. By $\sim 1\,\mathrm{Myr}$, an ionized region forms around the SMS and expels most of the remaining nearby gas, effectively terminating sustained gas accretion. After this point, the discrete jumps seen in Fig.~\ref{fig:3_new_MvsT} are produced primarily by mergers among 
SMSs, which dominate the residual mass growth at late times.

For $Z/Z_\odot = 10^{-3}$, the early growth is slower by about an order of magnitude because fine-structure line cooling reduces the initial Jeans mass. Consequently, the stellar mass reaches only $\sim10^{3}\,M_\odot$ by $10^{4}\,\mathrm{yr}$. At later times, however, the accretion rate increases to $\sim0.1$--$1\,M_\odot\,\mathrm{yr^{-1}}$ due to substantial inflow from the outer, hot region of the cloud, allowing the star to grow to $\sim10^{4}\,M_\odot$ by $\sim0.1\,\mathrm{Myr}$. Gas accretion then fades and largely ceases by $\sim 1\,\mathrm{Myr}$ after the initial protostar formation, after which SMS--SMS mergers again account for most of the remaining mass increase.

For $Z/Z_\odot = 10^{-2}$, the growth history is qualitatively different from the lower-metallicity cases. The mean accretion rate stays at $\sim10^{-3}\,M_\odot\,\mathrm{yr^{-1}}$ throughout the simulation, with large fluctuations associated with merger events. The stellar mass reaches $\sim2300\,M_\odot$ at $\sim0.3\,\mathrm{Myr}$, after which gas accretion is completely quenched by strong ionizing radiation. In the absence of stellar feedback, the mass could in principle continue to increase and approach $\sim10^{4}\,M_\odot$.

Finally, we note that the stellar evolution model adopted in \citet{Chon2025} is based on a simplified prescription. A more detailed stellar model could yield different internal structures and effective temperatures, which may in turn affect the strength of feedback and the final stellar mass. We discuss these model dependencies in Section~\ref{sec:discussion}.

For clarity, throughout Section~\ref{sec:results} we interpret “accretion” versus “collision” as smooth versus bursty segments of the adopted \(\dot M(t)\) histories, not as separate growth prescriptions inside the 1D stellar evolution calculation.

\subsection{Accretion and collision regulated growth of metal-enriched SMSs}
\begin{figure*}
    \centering
    \includegraphics[width=\linewidth]{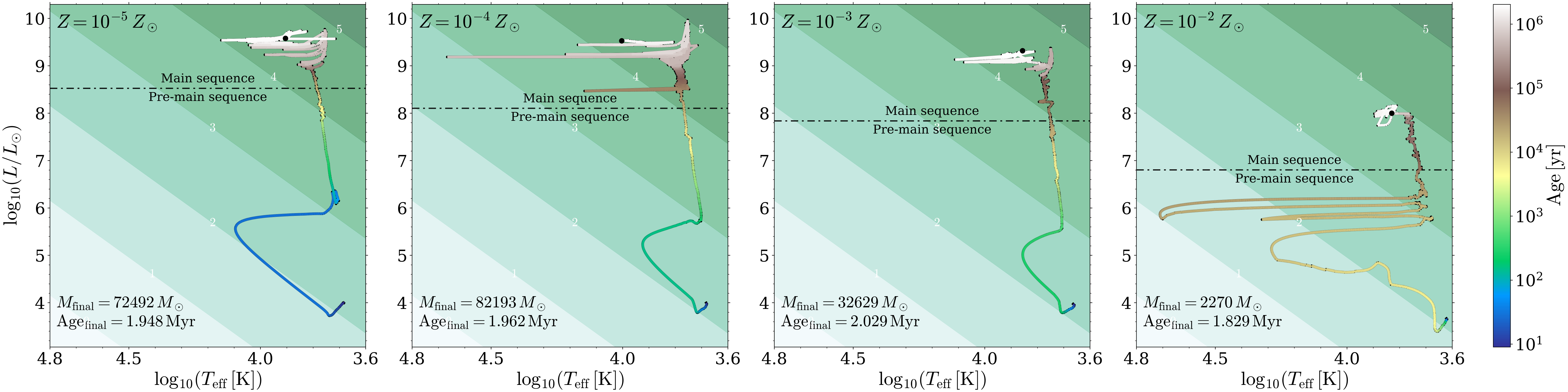}
    \caption{Hertzsprung–Russell diagrams for supermassive stars at four metallicities of $Z/Z_\odot=10^{-5}$, $10^{-4}$, $10^{-3}$, and $10^{-2}$ from left to right panels. Colourbar indicates stellar age increasing from blue to white. The dashed line separates pre–main-sequence and main-sequence phases.The diagonal background shading denotes \(\log_{10}(R/R_\odot)\) isoradius bands, with the numeric labels (1--5) marking the corresponding radius decades. Final masses and ages are $M_{\rm f}={7.25,8.22,3.26,0.23}\times10^{4}\,M_\odot$ and $t_{\rm f}={1.95,1.96,2.03,1.83}$ Myr.}
    \label{fig:1_HRD_Metallicities}
\end{figure*}

\begin{figure*}
    \centering
    \includegraphics[width=\linewidth]{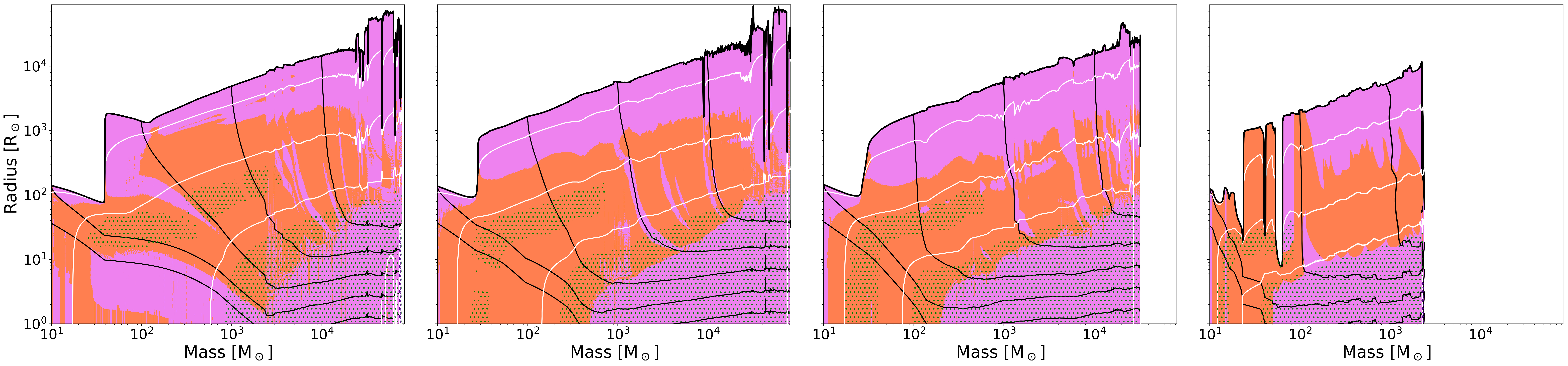}
    \caption{Kippenhahn diagrams (radius versus enclosed mass) for the same models but for the different metallicities of $Z/Z_\odot=10^{-5}$, $10^{-4}$, $10^{-3}$, and $10^{-2}$ from left to right panels. Lavender and coral regions denote convective and radiative zones, respectively; dotted areas mark active core-hydrogen burning. White contours are isotherms with $\log_{10}(T/{\rm K})=6$--8 (outermost $10^{6}\,$K and deepest $10^{8}\,$K) and black curves are iso-mass coordinates at enclosed masses of $1,10,10^{2},10^{3}, 10^{4}\,M_\odot$ with the set truncated by the final mass in each panel.}
    \label{fig:2_Kipp_Metallicities}
\end{figure*}

The four models in Fig.~\ref{fig:1_HRD_Metallicities} and Fig.~\ref{fig:2_Kipp_Metallicities} show how accretion and collisions shape the late pre–main-sequence and core–hydrogen-burning evolution of metal-enriched SMSs. Final masses decrease with increasing metallicity from \(7.2\times10^{4}\,M_\odot\) at \(Z/Z_\odot=10^{-5}\) to \(2.3\times10^{3}\,M_\odot\) at \(Z/Z_\odot=10^{-2}\), with luminosities dropping from \(\log(L/L_\odot)\simeq9.3\) to \(8.1\). All models end as inflated red supergiants with \(\log(T_{\rm eff}/{\rm K})\simeq3.8\)–4.2 (black dots in Fig.~\ref{fig:1_HRD_Metallicities}) and lifetimes of \(1.8\)–\(2.0\) Myr. Blueward excursions reflect short declines in the accretion rate. At \(Z\!<\!10^{-3}\,Z_\odot\), high inflow persists through core–H burning, producing repeated inflation–contraction cycles, while at \(Z=10^{-2}\,Z_\odot\) accretion fades after the pre–MS stage and the track settles near \(\log(T_{\rm eff}/{\rm K})\simeq3.8\). The Kippenhahn diagrams (Fig.~\ref{fig:2_Kipp_Metallicities}) reveal a convective core, a radiative intermediate zone, and a convective envelope sustained throughout the pre–MS and main sequence. The \(Z/Z_\odot=10^{-2}\) case becomes fully convective during core–H burning due to higher opacity and enhanced envelope mixing. We next examine the \(Z/Z_\odot=10^{-5}\) model in detail as the representative low-metallicity case.

\textit{Case \(Z/Z_\odot=10^{-5}\): evolution under sustained supercritical accretion} — The track starts at \(\log(T_{\rm eff}/{\rm K})=3.68\) and \(\log(L/L_\odot)=4.0\) (Fig.~\ref{fig:1_HRD_Metallicities}). The accretion rate remains above \(2\times10^{-2}\,M_\odot\,\mathrm{yr^{-1}}\) during the pre–MS phase, so the star evolves along the Hayashi line after the luminosity wave \citep{Hosokawa2010,Sakurai2015,Nandal2023,Nandal2024}, analogous to Pop~III SMS models \citep{Hosokawa2010,Woods2017}. Pre–MS evolution ends in the red at \(\log(T_{\rm eff}/{\rm K})=3.78\) and \(\log(L/L_\odot)=8.52\), where core–H burning begins. The accretion threshold itself decreases as burning proceeds, but at \(t=0.167\) Myr \(\dot M\) briefly dips below critical and the envelope contracts on its surface Kelvin–Helmholtz timescale \citep{Sakurai2015,Nandal2023}. A collision at \(t=0.183\) Myr then drives \(\dot M>10\,M_\odot\,\mathrm{yr^{-1}}\), inflates the envelope, and lowers \(\log(T_{\rm eff}/{\rm K})\) to 3.76 at the Hayashi limit. By \(t=0.60\) Myr the mean \(\dot M\) declines while collisions persist; the model is at \(\log(T_{\rm eff}/{\rm K})=3.97\), \(\log(L/L_\odot)=9.23\), and \(X_c=0.58\). At \(t=0.72\) Myr, with \(\log(T_{\rm eff}/{\rm K})=4.02\) and \(\log(L/L_\odot)=9.38\), growth is collision-dominated; at \(2.5\times10^{4}\,M_\odot\) the Kippenhahn diagram shows radius changes by orders of magnitude after each event (Fig.~\ref{fig:2_Kipp_Metallicities}). Each collision triggers expansion toward the Hayashi line on the envelope Kelvin–Helmholtz timescale, distinct from the global Kelvin–Helmholtz timescale, and blue–to–red loops occur because the interval between collisions exceeds the thermal relaxation time of the outer layers. When \(X_c=0.10\) the GR–instability criterion is met and the run stops, at \(M_{\rm final}=7.2492\times10^{4}\,M_\odot\) and \(t=1.948\) Myr.

\textit{Cases $Z/Z_\odot=10^{-4}$ and $Z/Z_\odot=10^{-3}$: transition from collision-dominated to accretion-limited growth} — Both models start their evolution at nearly the same position in the Hertzsprung–Russell diagram as the $Z/Z_\odot=10^{-5}$ case, with minor offsets due to higher initial metallicities.  During the pre–main-sequence phase, their accretion rates remain above the critical limit of $2\times10^{-2}\,M_\odot\,{\rm yr^{-1}}$, keeping them near the Hayashi line.  In the $Z/Z_\odot=10^{-4}$ model, $\dot{M}$ briefly drops below the limit at $\log(T_{\rm eff}/{\rm K})=3.77$ and $\log(L/L_\odot)=8.50$, then recovers but declines steadily afterward.  By $0.72$\,Myr, at $\log(T_{\rm eff}/{\rm K})=3.85$ and $\log(L/L_\odot)=9.14$, accretion-driven mass gain has nearly ceased (see second panel of Fig.~\ref{fig:1_HRD_Metallicities}).  The star contracts and becomes hotter, reaching $\log(T_{\rm eff}/{\rm K})=4.66$ at $0.77$\,Myr before a new series of collisions expand it back to the Hayashi limit by $0.81$\,Myr.  This is evident in the second panel of Fig.~\ref{fig:2_Kipp_Metallicities} at 4$\times 10^{4}$ M$_\odot$, where the radius changes from 2.5$\times 10^{2}$ R$_\odot$ to 2$\times 10^{4}$ R$_\odot$.  Continued collisions maintain the model around $\log(T_{\rm eff}/{\rm K})\approx4.0$ until the central hydrogen fraction falls to $X_{\rm c}=0.05$, when the general-relativistic instability criterion is met.  The final mass is $M_{\rm final}=8.2193\times10^{4}\,M_\odot$ at an age of $1.962$\,Myr, making this the most massive model in the sequence and the one that gains the largest fraction of its mass through collisions.  

In contrast, the $Z/Z_\odot=10^{-3}$ model experiences a decline in gas accretion much earlier during hydrogen burning.  The inflow rate falls below the critical value at $t=1.1$\,Myr, $\log(T_{\rm eff}/{\rm K})=3.82$, and $\log(L/L_\odot)=9.08$.  Collisions continue to occur but deliver smaller mass increments, and the envelope expansion following each burst is correspondingly weaker. The model reaches the end of core-helium burning at $t=2.029$\,Myr with $M_{\rm final}=3.2629\times10^{4}\,M_\odot$, without entering the GR-unstable regime.  The sharp decrease in final mass between the \( Z/Z_\odot = 10^{-4} \) and \( Z/Z_\odot = 10^{-3} \) cases arises from the metallicity-dependent cooling efficiency: higher metal content enhances fine-structure line cooling (from C~\textsc{i} and O~\textsc{ii}), thereby reducing the initial Jeans mass. This, in turn, leads to less massive and more gravitationally stable circumstellar disks around the SMS, resulting in fewer companion stars. Consequently, the \( Z/Z_\odot = 10^{-3} \) model accretes less efficiently, and collisions contribute only minor fractional mass increases, marking the transition from collision-dominated to accretion-limited growth.

\textit{Case $Z/Z_\odot=10^{-2}$: growth limited by short-lived accretion and early contraction} — In this highest metallicity model, stellar growth proceeds primarily through accretion, which contributes nearly equally during the pre–main-sequence and main-sequence stages.  The accretion rate varies strongly throughout the pre–main-sequence phase, as seen in Fig.~\ref{fig:1_HRD_Metallicities}, where $\log(T_{\rm eff}/{\rm K})$ oscillates between 3.67 and 4.71, and in Fig.~\ref{fig:2_Kipp_Metallicities}, where the stellar radius changes by more than two orders of magnitude.  Once the pre–main-sequence is completed, $\dot{M}$ rises above the critical limit, and the track settles at the Hayashi limit.  When accretion terminates, the model reaches the end of core-hydrogen burning.  After a brief phase of structural readjustment, core-helium burning begins and proceeds until $\log(T_{\rm eff}/{\rm K})=3.85$ and $\log(L/L_\odot)=8.00$.  The final mass and age are $M_{\rm final}=2.27\times10^{3}\,M_\odot$ and $t_{\rm final}=1.829$\,Myr, respectively, and the model does not encounter the GR instability.

\subsection{Impact of accretion variability on stellar structure}
\begin{figure}
    \centering
    \includegraphics[width=\linewidth]{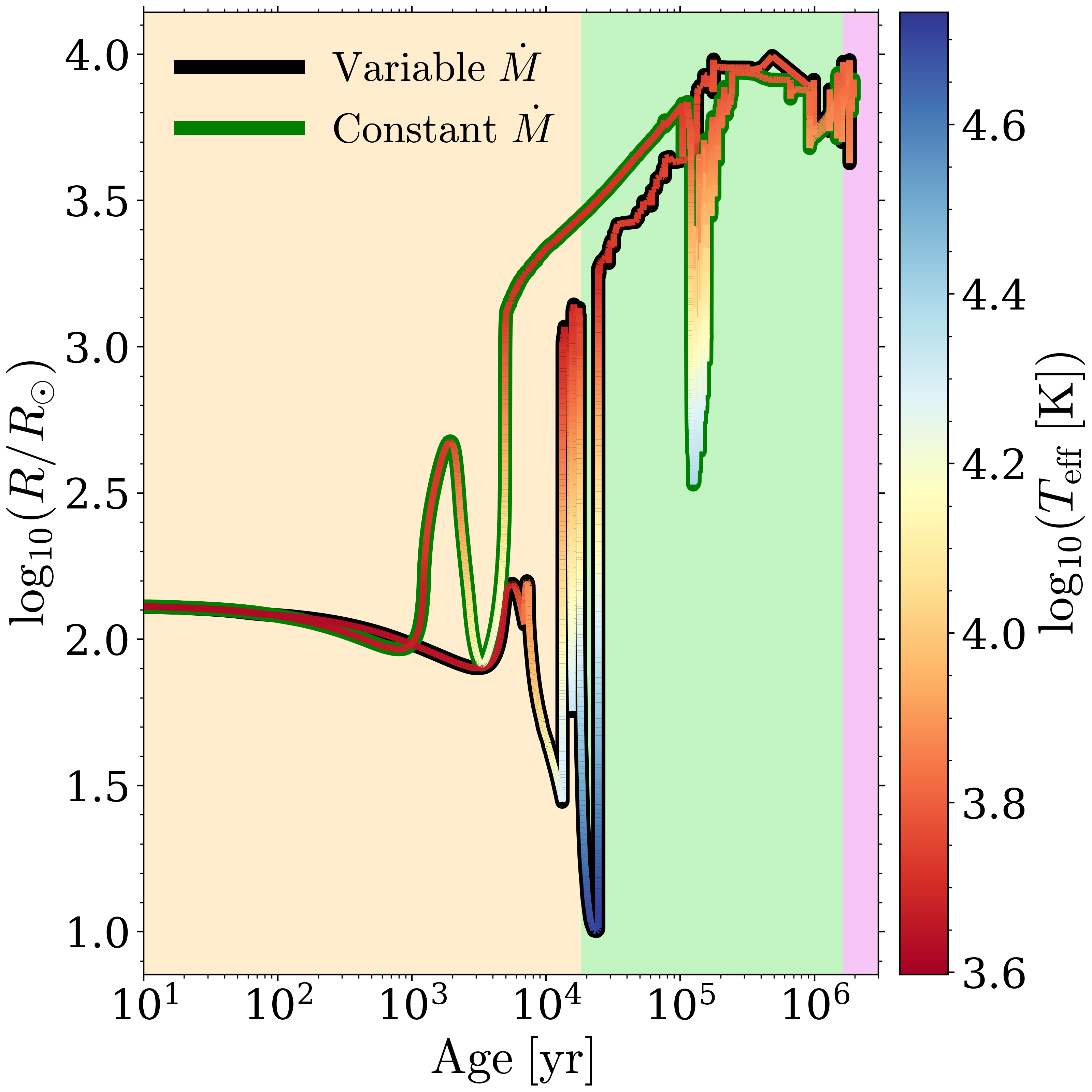}
    \caption{Radius evolution of two $Z/Z_\odot=10^{-2}~Z_\odot$ models reaching the same final mass ($\simeq2270,M_\odot$). The black-outlined track shows the variable accretion case; the brown-green track indicates the constant-rate model ($\dot{M}=9.321\times10^{-3},M_\odot,{\rm yr^{-1}}$). Colours denote effective temperature. Yellow, green, and purple shaded regions mark pre–main-sequence, core-hydrogen, and core-helium-burning phases, respectively.}
    \label{fig:3_R_vs_t_vs_Teff}
\end{figure}

We assess how the accretion history affects the structural evolution of the $Z/Z_\odot=10^{-2}$ model. Figure~\ref{fig:3_R_vs_t_vs_Teff} compares the radius evolution of two calculations that reach the same final mass of $\simeq2270\,M_\odot$: a model evolved with the time-dependent mass-growth history extracted from the three-dimensional simulation of \citet{Chon2025} (black-outlined track), and a model evolved at a constant accretion rate of $9.321\times10^{-3}\,M_\odot\,\mathrm{yr^{-1}}$ (brown-outlined track), equal to the time-averaged rate of the variable case. The two tracks overlap at early times but diverge once the accretion histories separate, with the constant-rate model undergoing a brief contraction to the ZAMS before re-inflating.


Both models evolve nearly identically during the first $10^{3}\,$yr, when the variable accretion rate happens to match the constant value. Thereafter, their behavior diverges depending on how $\dot{M}$ compares to the evolving critical rate, $\dot{M}_{\rm crit}(X_{\rm c},Z)$. In the constant-$\dot{M}$ case, the star initially moves toward the Hayashi limit but, since the imposed inflow falls below $\dot{M}_{\rm crit}$, it cannot maintain an inflated envelope and rapidly contracts, returning to the ZAMS at $9\times10^{4}\,$yr, where core-hydrogen burning begins. Over the subsequent $\sim10^{5}\,$yr, nuclear energy release inflates the envelope again and drives the star back toward the Hayashi limit. It then remains near the Hayashi limit until the end of core-helium burning, finishing at $\log(R/R_\odot)\simeq3.9$ and $\log(T_{\rm eff}/\mathrm{K})\simeq3.75$ at an age of $1.81\,\mathrm{Myr}$.

The variable-accretion model, in contrast, exhibits continuous radial oscillations driven by fluctuations in the inflow rate. Because its accretion rate stays near or above $\dot{M}_{\rm crit}$ during hydrogen burning, the star remains near the Hayashi limit throughout both the core-hydrogen and helium-burning phases. Despite these distinct pre--main-sequence and main-sequence histories, the two models converge to nearly identical final radii and effective temperatures. The variable-accretion model also finishes core-helium burning at $\log(R/R_\odot)\simeq3.9$ and $\log(T_{\rm eff}/\mathrm{K})\simeq3.75$, but at a slightly later age of $1.91\,\mathrm{Myr}$.

\begin{figure*}
    \centering
    \includegraphics[width=\linewidth]{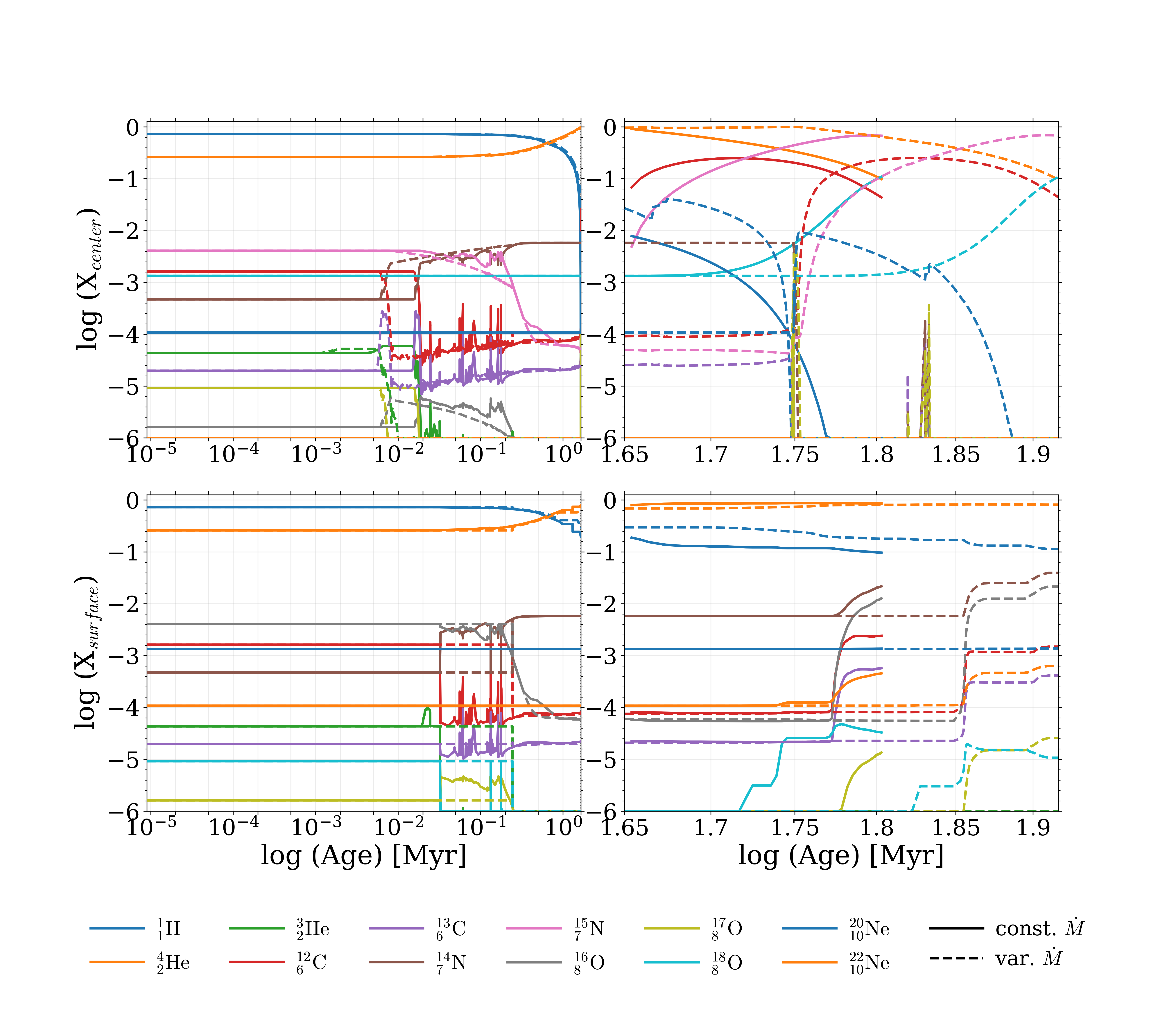}
    \caption{Central (top) and surface (bottom) mass fractions versus age for the two $Z/Z_\odot=10^{-2}~Z_\odot$ models. Solid lines show the constant accretion case ($\dot{M}=9.321\times10^{-3}\,M_\odot\,\mathrm{yr^{-1}}$); dashed lines show the variable accretion model based on 3D hydrodynamic histories. Plots in the left and right columns display abundances during pre–main-sequence + core-H and core-He burning, respectively. Multiple nuclear species are plotted; the discussion in the text focuses on H, He, $^{12}$C, $^{14}$N, and $^{16}$O.
}
    \label{fig:4_Central_surface_Abund}
\end{figure*}

Figure~\ref{fig:4_Central_surface_Abund} shows the time evolution of the central (top panels) and surface (bottom panels) abundances for the two $Z/Z_\odot=10^{-2}$ models discussed above: one accreting at variable rates (dashed lines) and the other at a constant rate of $9.321\times10^{-3}\,M_\odot\,{\rm yr^{-1}}$ (solid lines).  Apart from the differences in age, both models display remarkably similar chemical evolution, indicating that the internal transport of elements is only weakly affected by the accretion history once the stellar structure becomes largely convective.  

In the central regions (top-left panel of Fig.~\ref{fig:4_Central_surface_Abund}), we focus on H, He, $^{12}$C, $^{14}$N, and $^{16}$O.  On a logarithmic scale, the profiles of both models are nearly indistinguishable, since the fully convective cores efficiently homogenize the composition on the convective mixing timescale, $t_{\rm mix} \simeq \frac{R_{\rm conv}}{v_{\rm conv}}$, which in these models is on the order of a few years—much shorter than the nuclear timescale.  The only noticeable deviation occurs for $^{12}$C, with a difference of $\Delta X(^{12}{\rm C}) \simeq 10^{-5}$ in mass fraction between the two cases.  The top-right panel, corresponding to the core-helium-burning stage, shows the central abundances of $^{12}$C, $^{14}$N, $^{16}$O, and $^{4}$He to be effectively identical, confirming that nuclear processing proceeds at the same rate irrespective of the accretion history.

The bottom panels of Fig.~\ref{fig:4_Central_surface_Abund} trace the surface abundances during the pre–main-sequence and main-sequence phases (left) and during core-helium burning (right).  As discussed in Figs.~\ref{fig:1_HRD_Metallicities} and \ref{fig:2_Kipp_Metallicities}, the variable-accretion model undergoes large structural adjustments caused by episodic inflow, alternating between convective and radiative envelopes. Despite this structural variability, the transport of chemical species to the surface remains largely unaffected.  Up to an age of 1\,Myr, the surface hydrogen and helium mass fractions of the variable-accretion model are lower than in the constant-rate case by $\Delta X_{\rm H}\simeq0.1$ and $\Delta X_{^4{\rm He}}\simeq0.05$, respectively, while the CNO elements remain essentially identical.  This is consistent with the expectation that CNO-cycle processing during hydrogen burning contributes negligibly to global surface enrichment at this stage.  

During core-helium burning (bottom-right panel), the surface abundances converge further.  Hydrogen fractions are $X_{\rm H}=0.10$ for the variable case and $0.12$ for the constant case, while helium is $X_{\rm He}=0.90$ and $0.92$, respectively.  For the CNO elements, $^{12}$C reaches $4\times10^{-3}$ (constant) and $2\times10^{-3}$ (variable); $^{14}$N is $4\times10^{-2}$ (constant) and $6\times10^{-2}$ (variable); and $^{16}$O is $1.5\times10^{-2}$ (constant) and $3.5\times10^{-2}$ (variable).  These differences, all within a factor of $\lesssim2$, are negligible in the context of the overall chemical evolution.  The results demonstrate that by the end of core-helium burning, the signatures of the accretion history, whether constant or highly variabl are effectively erased.  Consequently, surface abundances alone cannot reveal whether a supermassive star experienced steady or bursty accretion.  This finding suggests that, for studies of the internal structure and chemical evolution of such objects, adopting a constant accretion rate provides an accurate approximation. Accretion during advanced burning phases could, in principle, produce minor differences, but these are unlikely to alter the overall structure. Including the effects of rotation would further enhance internal mixing and reduce any remaining disparities, leaving the two evolutionary paths virtually indistinguishable.

\subsection{Inefficiency of Collisional Mass Gain for Sustaining SMS Evolution}
\label{subsec:rejuvenation_delay}

To isolate the physical consequences of collisions on stellar structure, we adopt an idealized upper-limit treatment in which each collisional encounter deposits \emph{all} of its mass onto the growing supermassive star. No dynamical mass loss, radiative ablation, or nuclear burning triggered by impact heating is included. This represents a best-case scenario for collisional growth, designed to probe the extent to which successive mergers can rejuvenate the hydrogen-burning core and drive a runaway sequence of mass accumulation. In this framework, collisions are treated purely as instantaneous increments to the surface mass reservoir, allowing us to quantify the timescale and magnitude of subsequent structural adjustments within the star.

\begin{figure*}
    \centering
    \includegraphics[width=\linewidth]{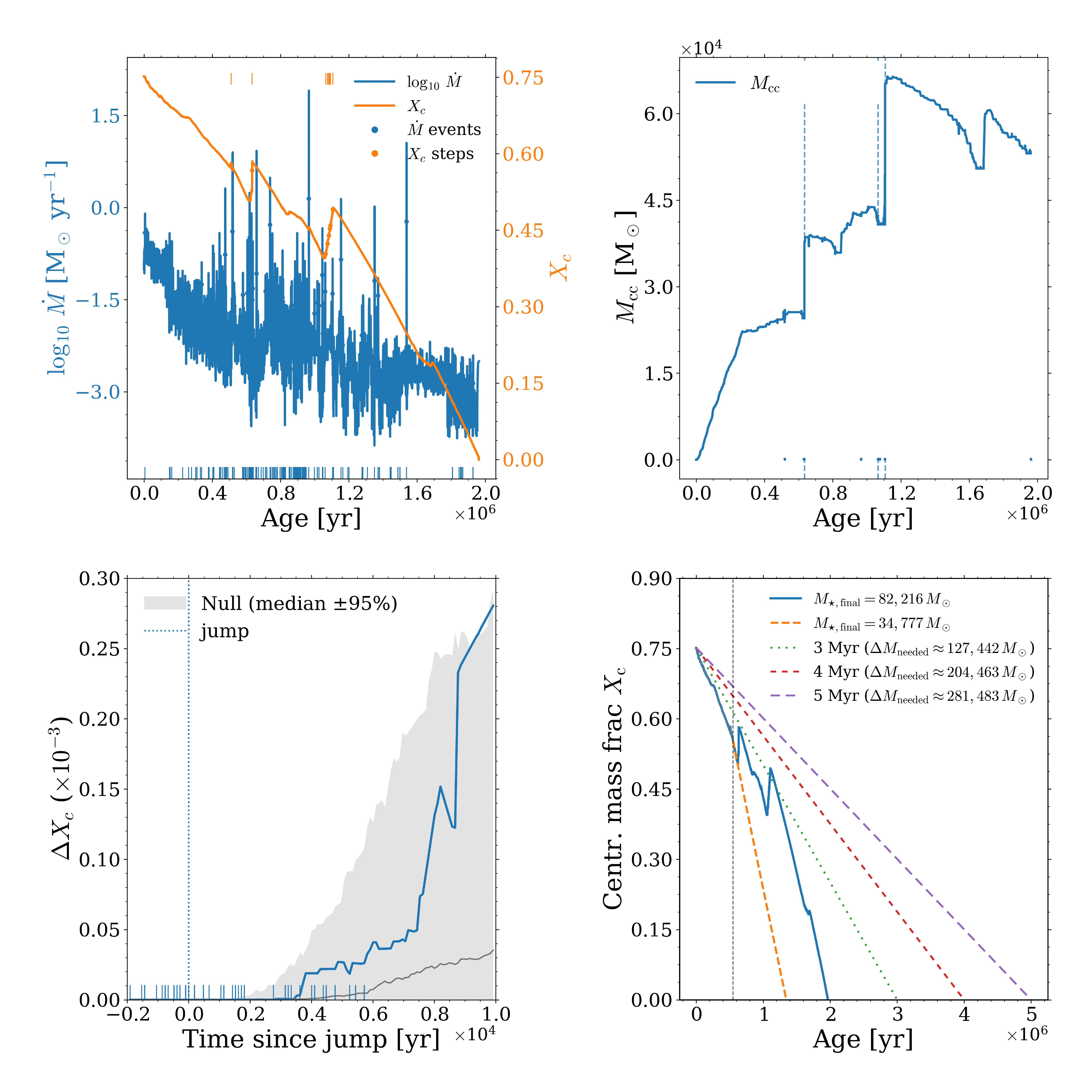}
    \caption{Collision diagnostics for the $Z=10^{-4}\,Z_\odot$ SMS.
    \textit{Top left:} Age on the $x$–axis with twin $y$–axes for $\log_{10}\dot M$ (blue) and $X_{\rm c}$ (orange). Filled markers flag detected accretion jumps and $X_{\rm c}$ steps.
    \textit{Top right:} Convective–core mass $M_{\rm cc}$ versus age with vertical dashed lines marking rejuvenation times.
    \textit{Bottom left:} Event–stacked $\Delta X_{\rm c}(t)$ around detected jumps showing the median signal (blue) against a permutation null band (median $\pm95\%$, grey). The grey solid line indicates the median. The $y$–axis is in units of $\times10^{-3}$.
    \textit{Bottom right:} $X_{\rm c}$ versus age for the fiducial collision+accretion track (solid blue) and a counterfactual run with removing the mass gain via collision after $t=0.55$ Myr (dashed orange). Legends report $M_{\star,\mathrm{final}}$ and thin lines indicate linear lifetime targets to 3, 4, and 5 Myr.}
    \label{fig:5_Collision_analysis}
\end{figure*}

Top left panel of Figure~\ref{fig:5_Collision_analysis} illustrates this behaviour for the $Z=10^{-4}\,Z_\odot$ model, where the time-dependent accretion rate (left $y$-axis) is overplotted with the evolution of the central hydrogen mass fraction $X_{\mathrm{c}}$ (right $y$-axis). The accretion history consists of two distinct events: long intervals of quiescent inflow with $\dot{M}\!\sim\!10^{-2}$--$10^{-1}\,M_\odot\,\mathrm{yr^{-1}}$, while also punctuated by sharp spikes exceeding $10\,M_\odot\,\mathrm{yr^{-1}}$, each corresponding to a collision event within the parent cluster. These bursts dominate the cumulative mass budget and represent merger-driven mass deposition episodes rather than smooth gas inflow. The amplitude and frequency of these spikes decline with age, reflecting the depletion of neighbouring massive stars available for collisions.

The overlaid $X_{\mathrm{c}}(t)$ curve shows that each major accretion spike is followed by a discrete upward fluctuation in the central hydrogen fraction. For example, at $t\simeq0.21$\,Myr and again near $t\simeq0.74$\,Myr, $\dot{M}$ rises by over two orders of magnitude within a few hundred years, and $X_{\mathrm{c}}$ subsequently increases by $\Delta X_{\mathrm{c}}\!\approx\!0.05$--$0.06$ relative to its pre-collision baseline. These increments occur after finite delays of several$\,\times\!10^{3}$\,yr, consistent with convective transport timescales from the envelope to the hydrogen-burning core. Between events, $X_{\mathrm{c}}$ declines steadily due to nuclear consumption, producing a characteristic sawtooth pattern that traces the cycle of depletion, collision, and rejuvenation. The repeated recovery of $X_{\mathrm{c}}$ confirms that collisions not only add mass but also replenish the hydrogen reservoir available for fusion, extending the effective main-sequence lifetime of the star.

Now we move to the top-right panel of Fig.~\ref{fig:5_Collision_analysis}, which tracks the convective–core mass as a function of time for the $Z=10^{-4}\,Z_\odot$ model. The trend is monotonic on long timescales with superposed steps that match with clusters of collision spikes in the accretion history. A prominent example occurs at $t\simeq1.07$~Myr, where the core mass rises from $4.45\times10^{4}\,M_\odot$ to $6.5\times10^{4}\,M_\odot$. This step follows the burst sequence in the top-left panel and lags the onset of the spikes by a few$\times10^{3}$~yr, consistent with the delay inferred there from the $X_{\rm c}$ response. The behaviour confirms that collisions do not merely add envelope mass, they also drive rapid inward transport that enlarges the hydrogen-burning core. As the star grows, the luminosity and nuclear consumption rate increase, so $X_{\rm c}$ continues to decline between events despite the transient core-mass gains. The panel therefore corroborates the top-left result: repeated mergers rejuvenate the core and extend burning locally, yet the global nuclear clock advances as expected under the rising energy output of the expanding SMS.

The bottom–left panel stacks the cumulative change in central hydrogen, $\Delta X_{\rm c}(t)$, around each detected accretion jump. The blue curve is the event–median cumulative response. The gray solid line is the permutation–null median from time–shuffled $dX_{\rm c}/dt$. The gray band is the 95\% central interval of that null. Before the jump ($t<0$) the blue curve coincides with the null median, so there is no pre–event drift. After the jump the response remains flat for $\sim10^{3}$\,yr, then separates from the null median by $t\!\gtrsim\!2\times10^{3}$\,yr and reaches $\simeq(2.5$–$2.7)\times10^{-4}$ by $10^{4}$\,yr (axis in units of $10^{-3}$), while the null median stays near $\simeq0.7\times10^{-4}$. The signal therefore shows a sustained positive offset relative to the null median and approaches the upper edge of the 95\% null envelope at late times. The implied transport lag of $10^{3}$–$4\times10^{3}$\,yr is consistent with convective mixing and agrees with the contemporaneous growth of the convective–core mass in the top–right panel.

Finally, the bottom-right panel of Fig.~\ref{fig:5_Collision_analysis} compares the central hydrogen fraction versus age for two $Z=10^{-4}\,Z_\odot$ tracks. The fiducial collision\,+\,accretion model is the solid blue line; the orange line recomputes the evolution from \(t=0.55\)\,Myr with the \emph{collisional} mass removed and only the smooth accretion history retained. The orange track ends at $M_{\rm f}=3.4777\times10^{4}\,M_\odot$ and $t_{\rm f}=1.35$\,Myr, whereas the fiducial model reaches $M_{\rm f}=8.219\times10^{4}\,M_\odot$ and $t_{\rm f}=1.942$\,Myr. Thus the blue–orange difference isolates the effect of collisions: adding \(\Delta M_{\rm coll}\simeq4.74\times10^{4}\,M_\odot\) extends the lifetime by \(\Delta t\simeq0.59\) Myr. This weak dependence is expected for radiation pressure dominated SMSs whose luminosities track the Eddington limit, $L\!\propto\!M$; the available fuel scales with $M X$, so the nuclear timescale $t_{\rm nuc}\!\sim\!(M X)/L$ depends only weakly on $M$ once $X_{\rm c}$ is set by mixing \citep{Woods2020, herr23a}. Collisions do rejuvenate $X_{\rm c}$ (top panels) but they also raise $L$ nearly proportionally, so the net gain in lifetime remains modest unless the core hydrogen fraction is substantially increased and maintained. A linear extrapolation of the measured slope between these two tracks implies that extending the lifetime to 3, 4, and 5\,Myr would require additional collisional mass of $\approx1.27\times10^{5}$, $2.04\times10^{5}$, and $2.81\times10^{5}\,M_\odot$, respectively; the true requirement is likely larger because $L(M)$ steepens with structural inflation and because $X_{\rm c}(t)$ is not strictly linear. In the idealized limit adopted here (100\% mass gained from collision, no impact heating, no collisional mass loss), these masses already strain plausibility for a runaway sequence: at $15\,M_\odot$ per impact, the 3\,Myr case implies $\sim8.5\times10^{3}$ mergers, far above typical collisional yields inferred for dense clusters even before including winds and dynamical ejections. We conclude that collisions can extend the lifetime but, by themselves, are an inefficient path to sustain SMS growth; efficient gas accretion remains necessary to reach and maintain supermassive masses.

\subsection{Critical accretion rate during core-hydrogen burning}
\label{subsec:crit_mdot}
In this subsection we define the \emph{critical accretion rate}, \(\dot M_{\rm crit}\), as the minimum steady mass-inflow rate required to keep a supermassive star on the Hayashi line (i.e., as a red–supergiant structure) during core-hydrogen burning. Previous works for zero‐metallicity models found \(\dot M_{\rm crit}\sim10^{-3}\)–\(10^{-2}\,M_\odot\,\rm yr^{-1}\) \citep{Omukai2003,Hosokawa_2010, Woods2017, Lionel2018} and more recently \(\dot M_{\rm crit}\simeq2.5\times10^{-2}\,M_\odot\,\rm yr^{-1}\) \citep{Nandal2023} during the pre–MS phase of Pop III stars. Here we extend the analysis to metal-enriched models and show that \(\dot M_{\rm crit}\) depends on both evolutionary phase and metallicity across \(Z/Z_\odot=10^{-4}\)–\(10^{-2}\).

\begin{figure*}
    \centering

    \includegraphics[width=\linewidth]{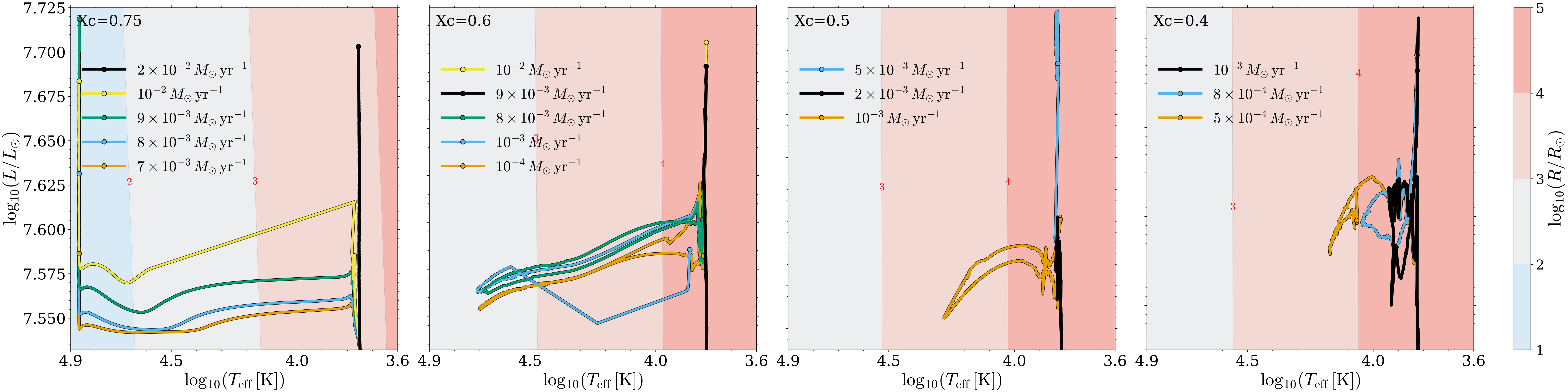}\\[2pt]
    \includegraphics[width=0.51\linewidth]{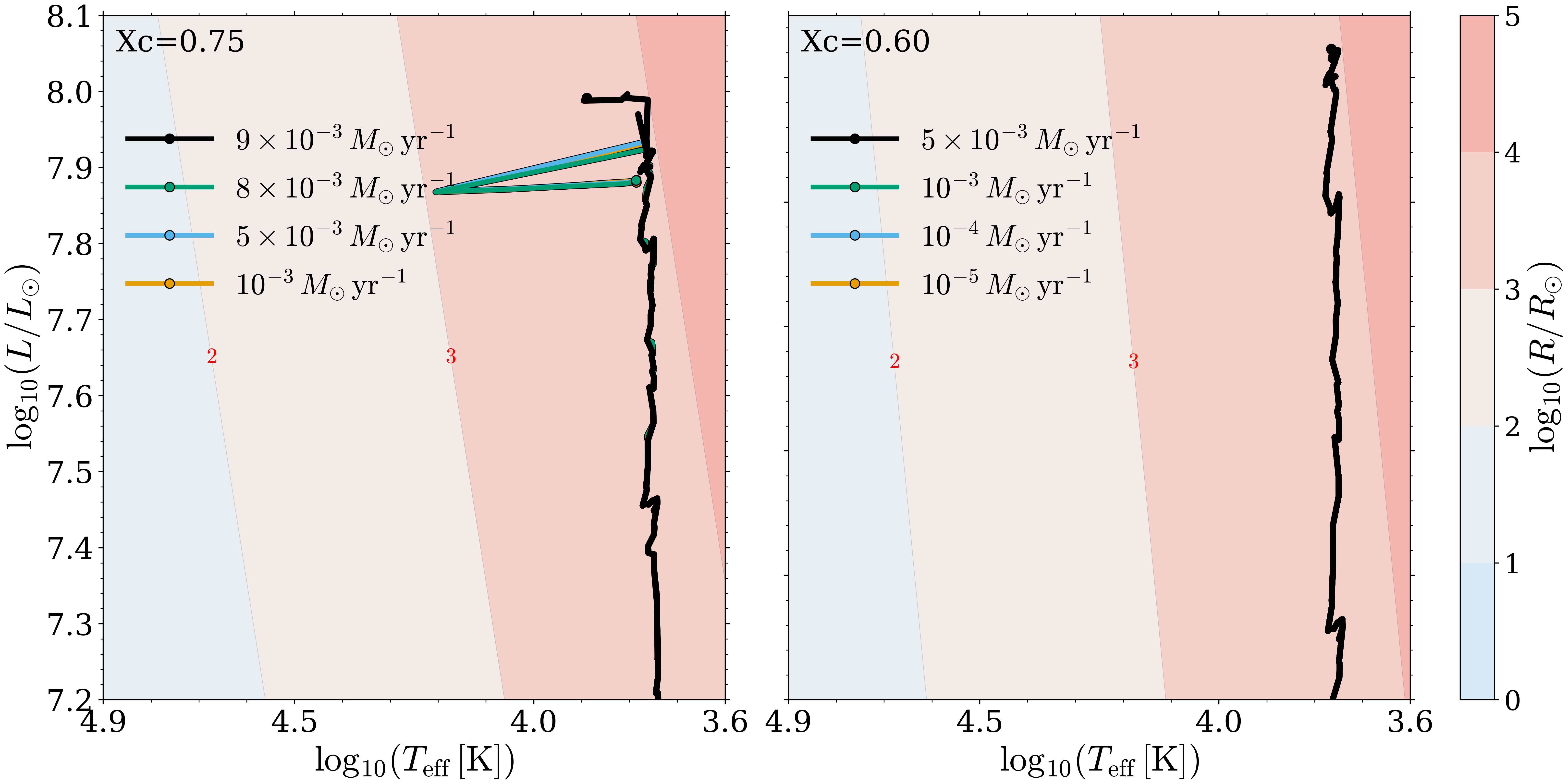}
    \caption{HR diagrams at fixed core–hydrogen fractions \(X_{\rm c}\) showing tracks computed at constant \(\dot M\). Top row: \(Z/Z_\odot=10^{-4}\) with \(X_{\rm c}=\{0.75,\,0.60,\,0.50,\,0.40\}\). Bottom row: \(Z/Z_\odot=10^{-2}\) with \(X_{\rm c}=\{0.75,\,0.60\}\). In each panel, the black track marks the case at \(\dot M_{\rm crit}\) that separates evolution staying on the Hayashi branch from blueward contraction at lower \(\dot M\). Background shading shows \(\log_{10}(R/R_\odot)\) isoradius bands with red numbers indicating the transition.}
    \label{fig:6_HRD_Critical_Mdot}
\end{figure*}

Figure~\ref{fig:6_HRD_Critical_Mdot} summarizes our constant-$\dot{M}$ experiments used to measure $\dot{M}_{\rm crit}$. For each metallicity, we restart the evolution from models at fixed central hydrogen mass fraction, $X_{\rm c}$, impose a constant accretion rate, and determine the threshold below which the star contracts and migrates blueward in the HR diagram.

\begin{figure}
    \centering
    \includegraphics[width=\linewidth]{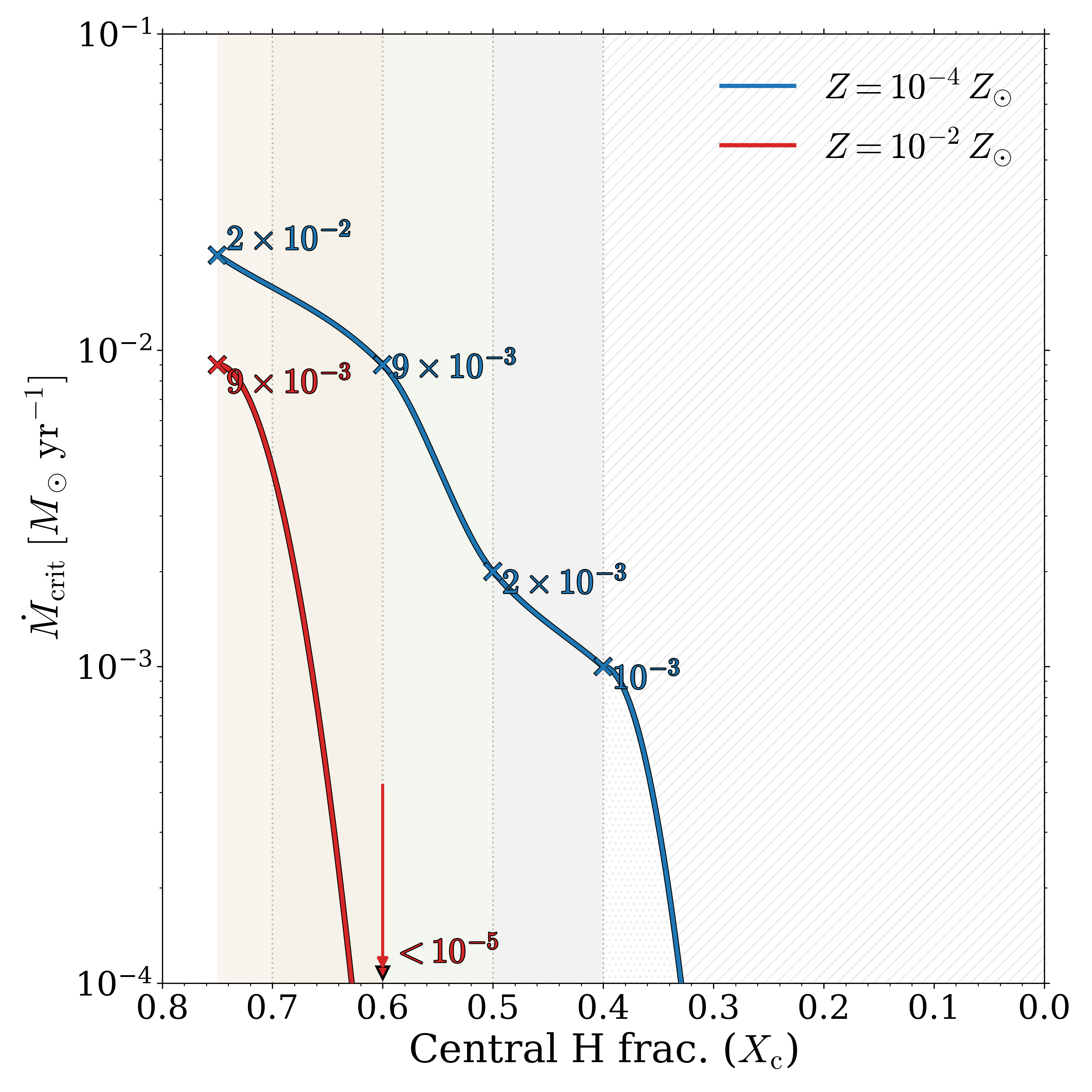}
    \caption{Critical accretion threshold $\dot{M}_{\rm crit}(X_{\rm c})$ during core--H burning for two metallicities, $Z/Z_\odot=10^{-4}$ (blue) and $Z/Z_\odot=10^{-2}$ (red). Crosses mark the $\dot{M}_{\rm crit}$ values measured from constant-$\dot{M}$ experiments at discrete $X_{\rm c}$, with the numerical values annotated on the plot. The downward arrow indicates an upper limit at $X_{\rm c}=0.60$ for $Z/Z_\odot=10^{-2}$, where no critical threshold is found within the explored $\dot{M}$ grid and the star remains inflated down to the lowest tested inflow rate. The blue and red curves are obtained by monotone piecewise-cubic Hermite interpolation (PCHIP) in $\log_{10}\dot{M}_{\rm crit}$ as a function of $X_{\rm c}$ between the labeled anchor points. Vertical bands mark the $X_{\rm c}$ intervals adopted in the HRD tests.}

    \label{fig:7_Critical_Mdot_vs_Xc}
\end{figure}

For $Z=10^{-4}\,Z_\odot$ (upper row of Fig.~\ref{fig:6_HRD_Critical_Mdot}), we begin with the model at the onset of core H burning ($X_{\rm c}\approx0.75$; first panel). At $\dot{M}=2\times10^{-2}\,M_\odot\,\mathrm{yr^{-1}}$ the track remains red for an additional $\sim3\times10^{5}\,$yr (chosen to exceed the surface thermal timescale of the inflated envelope), whereas lower rates trigger rapid contraction and blueward migration. Thus, at $X_{\rm c}\approx0.75$ we find $\dot{M}_{\rm crit}\simeq2\times10^{-2}\,M_\odot\,\mathrm{yr^{-1}}$, slightly below the pre--MS value of $2.5\times10^{-2}\,M_\odot\,\mathrm{yr^{-1}}$ reported by \citet{Nandal2023}. At later stages, the critical rate drops to $\dot{M}_{\rm crit}\simeq9\times10^{-3}$, $5\times10^{-3}$, and $1\times10^{-3}\,M_\odot\,\mathrm{yr^{-1}}$ at $X_{\rm c}=0.60$ (second panel), $0.50$ (third), and $0.40$ (fourth), respectively. With an updated set of tests at even later stages, we find that this decline continues beyond $X_{\rm c}\simeq0.40$, with $\dot{M}_{\rm crit}$ approaching $\sim10^{-4}\,M_\odot\,\mathrm{yr^{-1}}$ by $X_{\rm c}\simeq0.35$ (Fig.~\ref{fig:7_Critical_Mdot_vs_Xc}).

For $Z=10^{-2}\,Z_\odot$ (lower row of Fig.~\ref{fig:6_HRD_Critical_Mdot}), the corresponding experiments indicate a lower $\dot{M}_{\rm crit}$ than in the $Z=10^{-4}\,Z_\odot$ case. We tested the discrete set $\dot{M}=\{9\times10^{-3},\,5\times10^{-3},\,10^{-3},\,10^{-4},\,10^{-5}\}\,M_\odot\,\mathrm{yr^{-1}}$ at $X_{\rm c}\approx0.75$ and $\dot{M}=\{5\times10^{-3},\,10^{-3},\,10^{-4},\,10^{-5}\}\,M_\odot\,\mathrm{yr^{-1}}$ at $X_{\rm c}=0.60$. At $X_{\rm c}\approx0.75$ we find $\dot{M}_{\rm crit}\simeq9\times10^{-3}\,M_\odot\,\mathrm{yr^{-1}}$, a factor of $\simeq2$ lower than for $Z=10^{-4}\,Z_\odot$. At $X_{\rm c}=0.60$, the star remains on the Hayashi line for all tested inflow rates ($10^{-5}$--$5\times10^{-3}\,M_\odot\,\mathrm{yr^{-1}}$), so we can place only an upper limit of $\dot{M}_{\rm crit}\lesssim10^{-5}\,M_\odot\,\mathrm{yr^{-1}}$ at this stage.

At $Z=10^{-2}\,Z_\odot$ and $X_{\rm c}\approx0.75$ (lower-left panel of Fig.~\ref{fig:6_HRD_Critical_Mdot}), our constant-$\dot{M}$ tests show that the envelope stays inflated for most of the evolution even when $\dot{M}<\dot{M}_{\rm crit}$. The star may contract, but only briefly (a few $\times10^{3}\,$yr), reaching at most $\log_{10}(T_{\rm eff}/{\rm K})\simeq4.25$, and then re-expanding to the inflated Hayashi track.
In this sense, $\dot{M}_{\rm crit}$ at $X_{\rm c}\approx0.75$ still separates continuously inflated evolution from trajectories with a brief contraction. However, because the contraction is short and the temperature increase is modest, the radiative feedback remains weak and is unlikely to stop accretion.
A temporary dip below $\dot{M}_{\rm crit}$ therefore does not necessarily halt mass growth, which relaxes the conditions for forming and maintaining inflated SMSs at $Z\simeq10^{-2}\,Z_\odot$ compared to scenarios in which crossing $\dot{M}_{\rm crit}$ leads to a sustained ZAMS-like phase (see also Section~\ref{sec:discussion1}).

Figure~\ref{fig:7_Critical_Mdot_vs_Xc} condenses the HRD experiments into two physically distinct regimes. At early core-H burning ($X_{\rm c}\approx0.75$) a finite inflow is still required to prevent Kelvin--Helmholtz contraction of the inflated envelope, but the required threshold is metallicity dependent: $\dot{M}_{\rm crit}\simeq2\times10^{-2}\,M_\odot\,\mathrm{yr^{-1}}$ at $Z/Z_\odot=10^{-4}$ and $\dot{M}_{\rm crit}\simeq9\times10^{-3}\,M_\odot\,\mathrm{yr^{-1}}$ at $Z/Z_\odot=10^{-2}$. By mid core-H burning the behaviour diverges more strongly. At $Z/Z_\odot=10^{-4}$ the star still shows a measurable threshold down to $X_{\rm c}\simeq0.35$ where $\dot{M}_{\rm crit}\sim10^{-4}\,M_\odot\,\mathrm{yr^{-1}}$. At $Z/Z_\odot=10^{-2}$, the star remains on the Hayashi line at $X_{\rm c}=0.60$ even when the imposed inflow is reduced to $10^{-5}\,M_\odot\,\mathrm{yr^{-1}}$, so $\dot{M}_{\rm crit}$ ceases to be a bracketed quantity and we obtain only an upper limit. The inflated structure becomes self-sustaining earlier at higher $Z$, so external mass loading is no longer the controlling factor once core burning is established.

In summary, Fig.~\ref{fig:7_Critical_Mdot_vs_Xc} shows that $\dot{M}_{\rm crit}$ is not universal. At fixed evolutionary phase the threshold is lower at higher metallicity, with $\dot{M}_{\rm crit}\simeq9\times10^{-3}\,M_\odot\,\mathrm{yr}^{-1}$ at $X_{\rm c}\approx0.75$ for $Z/Z_\odot=10^{-2}$ compared to $\dot{M}_{\rm crit}\simeq2\times10^{-2}\,M_\odot\,\mathrm{yr}^{-1}$ for $Z/Z_\odot=10^{-4}$. Moreover, by $X_{\rm c}=0.60$ at $Z/Z_\odot=10^{-2}$ the star remains on the Hayashi line for all tested inflow rates down to $10^{-5}\,M_\odot\,\mathrm{yr}^{-1}$, so $\dot{M}_{\rm crit}$ is no longer bracketed and only an upper limit can be placed at this stage. For applications we therefore treat the labeled points in Fig.~\ref{fig:7_Critical_Mdot_vs_Xc} as an empirical prescription for $\dot{M}_{\rm crit}(X_{\rm c},Z)$ during core--H burning. Between the measured anchors we recommend interpolation in $\log_{10}\dot{M}_{\rm crit}$, while stages where no transition is found should be carried as upper limits rather than extrapolated as a unique threshold.

Physically, the reduction of \(\dot M_{\rm crit}\) during core-hydrogen burning arises because the core’s nuclear energy output inflates the envelope and delays the Kelvin–Helmholtz contraction of the surface layers, making it easier for lower accretion rates to maintain the red–supergiant structure. In other words, the condition \(\tau_{\rm accr}\lesssim\tau_{\rm KH,surf}\) is satisfied at lower \(\dot M\) once the star can sustain an extended envelope without relying on external mass-loading to maintain its high-entropy outer layers. This argument is consistent with analytic treatments for SMSs where at high accretion the star resides on the Hayashi limit with \(L\propto M\) and \(R\propto M^{1/2}\) \citep{Lionel2021}.

The metallicity dependence follows from how \(Z\) modifies both envelope support and the onset of efficient H burning. The envelope inflation of luminous stars is controlled by the local Eddington factor \(\Gamma=\kappa L/(4\pi cGM)\). At higher \(Z\), the Rosseland mean opacity \(\kappa\) is larger and opacity peaks become stronger, which raises \(\Gamma\) in the envelope and promotes inflation without needing any contrubution from the stellar interior. Detailed stellar-structure studies show that opacity peaks, including the iron bump, play a major role in setting the degree of envelope inflation and that the extent of inflation depends on metallicity through \(\kappa(Z)\) \citep{Sanyal2017}. In parallel, metal enrichment supplies CNO catalysts from the start of core-H burning, which reduces the degree of contraction required to reach a burning configuration that can maintain the near-Eddington luminosity. In the supergiant-protostar context, metal enrichment has been shown to lower the central temperature required for H burning and to introduce an additional opacity feature, both of which favour earlier envelope inflation \citep{Inayoshi2013}. In our models this combination manifests as a lower ZAMS normalization of \(\dot M_{\rm crit}\) at \(Z=10^{-2}\,Z_\odot\) (\(9\times10^{-3}\) rather than \(2\times10^{-2}\,M_\odot\,\rm yr^{-1}\) at \(X_{\rm c}\approx0.75\)) and as a rapid transition to a regime where we can only place an upper limit on \(\dot M_{\rm crit}\) by \(X_{\rm c}=0.60\).

\section{Discussion}\label{sec:discussion}

\subsection{Comparing the physics of stellar evolution} \label{sec:discussion1}
 
\begin{figure}
    \centering
    \includegraphics[width=\linewidth]{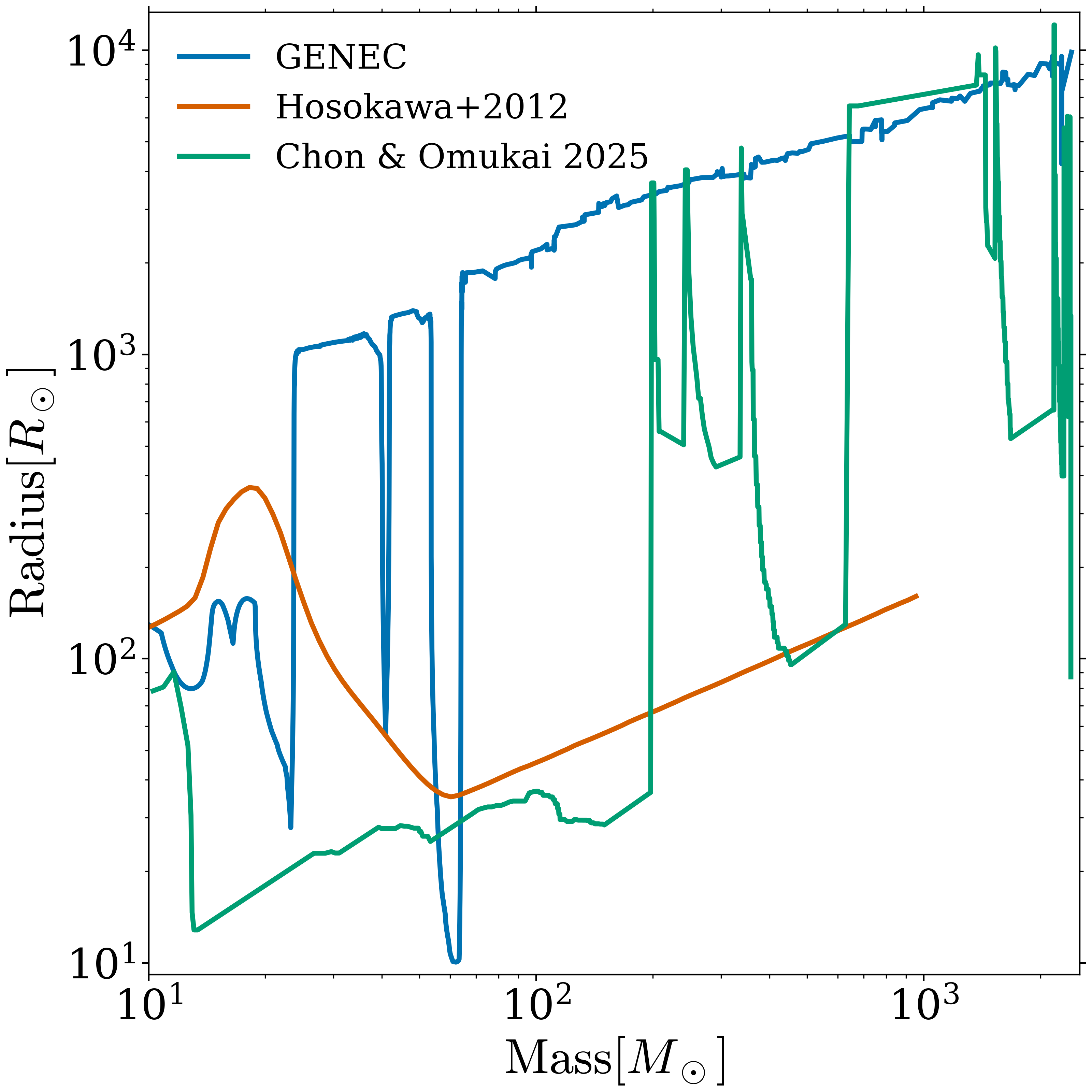}
    \caption{Radius--mass evolution at \(Z/Z_\odot=10^{-2}\) comparing three stellar-evolution treatments. Blue: our full-structure GENEC model evolved with the time-variable mass-delivery history including bursty collision-driven spikes. Orange: the constant-\(\dot M\) radius fits of \citet{Hosokawa2009}. Green: the framework of \citet{Chon2025}, which interpolates the \citet{Hosokawa2009} tables and switches to the supergiant prescription when \(\dot M\gtrsim2\times10^{-2}\,M_\odot\,\mathrm{yr^{-1}}\) following \citet{Sakurai2015}. Axes are shown in log scale.}
    \label{fig:R_vs_M_Z1e2}
\end{figure}

Figure~\ref{fig:R_vs_M_Z1e2} compares the radius versus mass evolution at \(Z/Z_\odot=10^{-2}\) for three models: GENEC with time-variable inflow plus collision spikes (blue), the constant-inflow fits of \citet[][orange]{Hosokawa2009}, and the 3D framework of \citet[][green]{Chon2025} that interpolates \citeauthor{Hosokawa2009} and activates surface inflation when \(\dot M \gtrsim 2\times10^{-2}\,M_\odot\,\mathrm{yr}^{-1}\) following \citet{Sakurai2015}. GENEC radii exceed the \citeauthor{Hosokawa2009} curve by \(\sim1\)–\(1.5\) dex over most of the range; at \(M\simeq2\times10^{3}\,M_\odot\) we find \(R\simeq3\times10^{4}\,R_\odot\) versus \(\lesssim10^{3}\,R_\odot\), and near \(M\simeq10^{4}\,M_\odot\) GENEC remains at a few\(\times10^{4}\,R_\odot\) while the fit stays below a few\(\times10^{3}\,R_\odot\). The \citet{Chon2025} track shows brief inflations at high \(\dot M\) yet stays bounded by the \citeauthor{Hosokawa2009} interpolation.

The residual offset arises from envelope entropy retention and mechanical work in our full-structure evolution. We assume cold disc accretion; freshly accreted gas sets a high-entropy surface, convection mixes this inward, the interior contracts on a Kelvin–Helmholtz timescale, and the outer layers expand by \(P\,\mathrm dV\) work. After core–H ignition the nuclear luminosity supports the inflated envelope, so the critical rate \(\dot M_{\rm crit}(X_c)\) decreases as \(X_c\) falls (Fig.~\ref{fig:7_Critical_Mdot_vs_Xc}). Our measured \(\dot M(t)\) stays above this moving threshold for long phases, which sustains \(R\sim10^{4\!-\!5}\,R_\odot\). A radius excess also appears at \(\dot M=9\times10^{-3}\,M_\odot\,\mathrm{yr}^{-1}\) because GENEC loses less entropy from the surface layers and permits more expansion before radiative leakage than the \citeauthor{Hosokawa2009} tables at the same \(\dot M\). The mechanism mirrors the drift of classical massive stars toward the red-supergiant branch \citet{Ekstrom2012, Nandal2023}; in supermassive stars it acts already during core–H burning due to large envelope mass and strong radiation pressure.

Our results open up the possibility that SMSs with masses up to \(10^4~M_\odot\) can form in environments with metallicities of \(10^{-2}~Z_\odot\), typical of those found in globular clusters \citep{Puzia2005, Beasley2019}. \citet{Chon2025} found that the mass of an SMS can reach up to \(2\times10^{4}~M_\odot\) when radiation feedback effects are neglected. Our simulations show that the supergiant phase persists into the later stages and reduces the emissivity of the ionizing photons, where the stellar mass exceeds \(100\)--\(1000~M_\odot\), which will allow the central star to grow efficiently \citep{Sakurai2015}. The weaker radiative feedback primarily results from the lower critical accretion rate required to maintain the supergiant phase at a low central hydrogen fraction (\(X_\mathrm{c}\)). Additional differences in energy transport within the radiative envelope further sustain efficient accretion. To fully assess the feasibility of SMS formation in globular cluster-forming environments, self-consistent stellar evolution calculations coupled with three-dimensional hydrodynamics are required \citep[e.g.,][]{Hosokawa2016, Sugimura2023}. Finally, effects of mass loss become increasingly more important with higher metallicities. The survivability of such red, near-Eddington SMSs (i.e. the stars remaining supermassive in presence of mass loss) after the post-growth phase must be studied via new and existing mass loss prescriptions in future stellar evolution calculations. 

In the literature, some studies have shown that SMSs can undergo energetic explosions rather than collapsing into black holes.  
In the primordial case, explosive helium burning coupled with GR instability may trigger such explosions \citep{Chen2014, Nagele2022}.  
The predicted mass range for these events is very narrow, around \(M = (5.55\!-\!5.60)\times10^{4}\,M_\odot\) in \citet{Chen2014}, but was later extended to a smaller mass range of \(M = (2.6\!-\!3.0)\times10^{5}\,M_\odot\) in \citet{Nagele2022}.  
In our calculations, the stellar masses lie close to these ranges, yet we still find that the stars collapse into BHs.  
This is consistent with the narrowness of the explosion window and the correspondingly small likelihood of such events.  
Metal enrichment facilitates explosions in SMSs with stellar masses around \(10^{5}\,M_\odot\) \citep{Fuller1986, Nagele2023, Nagele2024}.  
\citet{Nagele2023} and \citet{Nagele2024} followed the evolution of metal-enriched, rapidly accreting SMSs with \([Z/H] \gtrsim -1\) and demonstrated that nuclear burning can halt GR-induced collapse, leading to pulsations or full-scale explosions.  
In this work, we focus on lower metallicity regimes, below \(Z \sim 10^{-2}\,Z_\odot\), and do not find any signatures of explosion.  
We attribute this difference to the lower metallicity adopted here compared to that in \citet{Nagele2023}.  
This result is consistent with \citet{Fuller1986}, who found that SMS explosions do not occur for \(Z \lesssim 0.2\,Z_\odot\).  
We also note that our models include additional physical effects, such as stochastic collisional spikes, the measured decline of \(\dot{M}_{\mathrm{crit}}(X_{\mathrm{c}})\), and lifetime budgeting under variable mass supply.  
\citet{Nagele2023} and \citet{Nagele2024} did not quantify the collision-driven rejuvenation delays or the phase-dependent variations in \(\dot{M}_{\mathrm{crit}}\), which may lead to differences in the final stellar fate.  
Nevertheless, our results are consistent with theirs in showing that near-Eddington stellar interiors are only weakly sensitive (in lifetime) to additional mass accretion.  
In this sense, our work extends theirs from exploring the fates of SMSs at high \(Z\) and steady \(\dot{M}\) to investigating growth efficiency and stability thresholds at finite \(Z\) under bursty accretion histories.

\subsection{Globular clusters and collision dominated growth}
\citet{Gieles2018} proposed that SMSs with \(\gtrsim10^{3}\,M_\odot\) can form in globular clusters through repeated stellar collisions.
The foundation of their model is the continuous rejuvenation of the central star via successive mergers.  
Their “conveyor-belt’’ scenario suggests that, in dense clusters, repeated mergers can maintain an inflated, long-lived SMS as long as the supply of colliding stars persists.  
Although stellar winds limit mass growth, the extended stellar lifetime in this picture allows the star to reach supermassive scales.  
Our event-triggered analysis, however, indicates that collisions alone extend the stellar lifetime only modestly once the star approaches the Eddington limit.  
Because \(L \propto M\) and \(t_{\mathrm{nuc}} \sim (M X)/L\), the influence of additional mass on the lifetime becomes weak; thousands of mergers would be required to produce multi-Myr extensions at \(Z = 10^{-4}\,Z_\odot\).  
Thus, while we agree that collisions can rejuvenate the central hydrogen fraction (\(X_{\mathrm{c}}\)), we find that they do \emph{not} sustain indefinite stellar growth (via the "conveyor belt" scenario)  without continuous gas accretion. A higher collision rate or enhanced gas accretion, as found in \citet{Chon2025}, would be required to form SMSs in globular clusters.

While the mass-loss rates from stellar winds and collisions remain highly uncertain, they play an important role in linking the chemical signatures observed in globular clusters to the formation of SMSs \citep{Bastian2018}.  
\citet{Ramirez-Galeano2025} showed that, at most, about 10\% of the stellar mass would be lost through collisions. To accurately evaluate the chemical yields produced during such mergers, additional mixing of heavy elements synthesized deep within the stellar interior must be considered. Stellar rotation is a strong candidate for driving this internal chemical mixing \citep{deMink2009}, thereby providing a physical connection between the observed abundance patterns in globular clusters and the formation of SMSs. A self-consistent treatment of stellar mass growth, mass loss, and rotational evolution will be required to fully uncover the chemical composition of globular clusters formed through SMS evolution.

The formation of SMSs is also expected in galactic centers, where metal enrichment is likely to be significant \citep[e.g.,][]{Mayer2015}.  
\citet{Mayer2010} showed that violent gas accumulation can occur during major mergers of massive galaxies, leading to the formation of a supermassive disk at the galactic center. 
\citet{Zwick2023} analyzed the stability of such supermassive gas disks formed after galaxy mergers and found that they exhibit structures similar to those of supermassive stars due to rapid inflow.  
\citet{Zwick2025} further demonstrated that these systems could produce spectra resembling those of the recently observed Little Red Dots (LRDs) detected by JWST \citep[e.g.][]{Maiolino2024, Ubler2024, Harikane2025}.  
However, their study primarily focuses on large-scale gas inflow from galactic to stellar scales, rather than on the internal stellar physics.  
Our results therefore complement theirs: we model the internal response of metal-enriched SMSs under time-dependent accretion.  
We find that, once hydrogen burning ignites, the accretion rate required to remain on the Hayashi line decreases by more than an order of magnitude.  
This reduction facilitates the maintenance of SMS structure and promotes continued inflow, owing to the larger stellar radii, which in turn increases the number density of supermassive disks formed through galaxy mergers.

\subsection{Approximate mass-loss estimates at finite metallicity}

To obtain order-of-magnitude wind losses, we performed two tests at mid core–H burning for the $Z=10^{-4}\,Z_\odot$ model, when the stellar mass reached $7.3973\times10^{4}\,M_\odot$. Gas accretion and collisions were disabled, and the star was evolved with either the \citet{DeJager1988} or \citet{Vink2001} mass-loss prescriptions until H exhaustion.

With \citet{DeJager1988}, the final mass was $2.3421\times10^{4}\,M_\odot$, implying a loss of $\sim5.1\times10^{4}\,M_\odot$. This value is likely overestimated since the prescription was derived for Galactic supergiants near solar $Z$, lacks metallicity dependence, and is extrapolated to regimes where line driving and dust formation are inefficient. Using \citet{Vink2001} yields $M_{\rm f}=6.2348\times10^{4}\,M_\odot$, corresponding to a loss of $\sim1.2\times10^{4}\,M_\odot$. This wind becomes effective only during the short hot phases when the star contracts toward the ZAMS and its effective temperature exceeds $\sim2.5\times10^{4}\,$K, following episodes where the accretion rate falls below the critical threshold. During these O and B-type stages, line-driven winds can briefly operate, but the prescription itself remains untested for such extreme luminosities and low metallicities.

These tests therefore provide only first-order bounds: \citet{DeJager1988} gives a generous upper limit, while \citet{Vink2001} gives a conservative lower limit. No existing wind prescription is calibrated for supermassive, low-$Z$ stars, and a dedicated follow-up study will be needed to quantify mass loss and its feedback in this regime.

\section{Summary and Conclusion}\label{sec:conclusion}

This study presents the first self-consistent grid of metal-enriched supermassive star (SMS) models evolved with time-dependent accretion and idealized collisional mass gain using the \gva\ code.  We explored metallicities from $Z/Z_\odot=10^{-5}$–$10^{-2}$ and followed the stellar evolution through the main-sequence and, where applicable, to the onset of general-relativistic instability.  Our analysis isolates the structural and evolutionary response of SMSs to variable inflow, quantifies the critical accretion rate during core hydrogen burning, and evaluates whether collisions alone can sustain runaway growth.  The main findings are summarized below.

\begin{itemize}
    \item \textbf{Accretion and collision regulated growth.}  
    Across all metallicities, stars remain inflated near the Hayashi line while $\dot{M}\!>\!\dot{M}_{\rm crit}\!\simeq\!2\times10^{-2}\,M_\odot\,{\rm yr^{-1}}$.  
    Increasing $Z$ reduces the final mass from $7.2\times10^{4}$ to $2.3\times10^{3}\,M_\odot$ and suppresses collisional efficiency.  
    The transition from collision-dominated to accretion-limited growth occurs between $Z=10^{-4}$ and $10^{-3}\,Z_\odot$.

    \item \textbf{Impact of accretion variability.}  
    Variable accretion histories drive radial oscillations but leave the long-term structural and chemical evolution unchanged.  
    Constant-rate models reproduce identical final radii, lifetimes, and surface abundances, showing that the global evolution depends mainly on the mean $\dot{M}$, not its temporal variability.

    \item \textbf{Collision-induced rejuvenation.}  
    Collisions rejuvenate the hydrogen-burning core by $\Delta X_{\rm c}\!\simeq\!0.05$–$0.06$ with a mixing delay of a few$\times10^{3}$ yr, consistent with convective timescales.  
    The lifetime increase remains small: an additional $4.7\times10^{4}\,M_\odot$ of collisional mass extends the main sequence by only $\sim0.6$ Myr.

    \item \textbf{Critical accretion rate evolution.}  
    During core-H burning, $\dot{M}_{\rm crit}$ declines from $2\times10^{-2}$ to $10^{-3}\,M_\odot\,{\rm yr^{-1}}$ as $X_{\rm c}$ decreases from 0.75 to 0.40.  
    We additionally find a metallicity dependence, with \(\dot M_{\rm crit}\) lower at higher \(Z\) (e.g., \(9\times10^{-3}\,M_\odot\,{\rm yr^{-1}}\) at \(X_{\rm c}\approx0.75\) for \(Z=10^{-2}\,Z_\odot\)). Nuclear energy released in the core inflates the envelope and delays contraction of the surface layers, allowing the star to remain red at progressively lower inflow rates.

    \item \textbf{Inefficiency of runaway collisions.}  
    Even under 100\% retention efficiency, sustaining growth beyond $\sim2$ Myr would require thousands of mergers with massive stars, far exceeding plausible cluster encounter rates.  
    Collisions extend lifetimes but cannot replace high gas inflow; gas accretion remains the dominant channel for forming and maintaining metal-enriched SMSs.
    \item \textbf{A channel for SMS formation at higher metallicity.}  
    For \( Z/Z_\odot = 10^{-2} \), the star follows the Hayashi line until the end of core He burning, maintaining a cool envelope with negligible radiative feedback. This suggests a viable pathway for SMS formation even at metallicities as high as \( 10^{-2}\,Z_\odot \),  which are typical of globular cluster environments.
\end{itemize}

Future work will extend these models beyond core-hydrogen exhaustion into the late burning stages, ending in the production of pre-supernova yields for various metallicities.  We also tested existing mass-loss prescriptions, but the derived values remain highly uncertain, highlighting the need for a dedicated follow-up study focused on wind physics in the supermassive regime. Including rotation and magnetic transport will quantify angular-momentum retention and its effect on convective mixing and mass loss.  The resulting final masses and composition profiles will be coupled to three-dimensional hydrodynamic simulations such as \textsc{Arepo} to track the collapse and accretion of the remnant black hole, linking stellar-scale evolution to the formation and early growth of supermassive black holes in the high-redshift Universe. 
Zero–metallicity supermassive stars have been proposed as progenitors of Little Red Dots \citep[e.g.,][]{Naidu2025, deGraaff2025, Nandal2025c, Zwick2025}; we will test this hypothesis for metal-enriched SMSs in a forthcoming study.

\begin{acknowledgments}
DN was supported by the Swiss National Science Fund (SNSF) Postdoctoral Fellowship, grant number: P500-2235464. The authors thank Dr. K. Omukai for the detailed and fruitful discussions on critical accretion rates and the properites of stellar envelopes. We would also like to the thank the anonymous referee for helping us improve the preliminary versions of this manuscript. 
\end{acknowledgments}

\bibliography{sample7}{}
\bibliographystyle{aasjournalv7}



\end{document}